\documentclass[runningheads]{llncs}
\usepackage{graphicx}

\usepackage{comment}
\usepackage{amsmath,amssymb} %
\usepackage{color}
\usepackage[labelsep=period]{caption}
\usepackage{epsfig}
\usepackage{mathtools}
\usepackage{multirow}
\usepackage{authblk}
\usepackage[symbol]{footmisc}
\usepackage{multirow}
\usepackage{ctable} 
\usepackage[position=top]{subfig}
\usepackage{float}
\usepackage[normalem]{ulem}
\useunder{\uline}{\ul}{}

\usepackage{multirow}
\usepackage{subfig}
\usepackage{adjustbox}
\def\bal#1\eal{\begin{align}#1\end{align}} %

\def\m{\mathbf}

\newcommand {\bbmtx}{\begin{bmatrix}} %
\newcommand {\ebmtx}{\end{bmatrix}} %

\newcommand{\etal}{\textit{et al. }}
\newcommand{\rulesep}{\unskip\ \vrule\ }

\begin{document}
\pagestyle{headings}
\mainmatter

\title{Microscopy Image Restoration with Deep Wiener-Kolmogorov Filters} %

\titlerunning{}

\author{Valeriya Pronina \inst{1} \and
Filippos Kokkinos\inst{2} \and
Dmitry V. Dylov\inst{1}\and Stamatios Lefkimmiatis\inst{3}}

\authorrunning{V.Pronina et al.}
\institute{Skolkovo Institute of Science and Technology, Moscow, Russia \and
University College London \and Q.bio Inc.}
\maketitle

\begin{abstract}
    Microscopy is a powerful visualization tool in biology, enabling the study of cells, tissues, and the fundamental biological processes; yet, the observed images typically suffer from blur and background noise. In this work, we propose a unifying framework of algorithms for Gaussian image deblurring and denoising. These algorithms are based on deep learning techniques for the design of learnable regularizers integrated into the Wiener-Kolmogorov filter. Our extensive experimentation line showcases that the proposed approach achieves a superior quality of image reconstruction and surpasses the solutions that rely either on deep learning or on optimization schemes alone. 
    Augmented with the variance stabilizing transformation, the proposed reconstruction pipeline can also be successfully applied to the problem of Poisson image deblurring, surpassing the state-of-the-art methods. 
    Moreover, several variants of the proposed framework demonstrate competitive performance at low computational complexity, which is of high importance for real-time imaging applications.
    \keywords{deblurring, denoising, learnable regularizers, microscopy deblurring, Wiener filter.}
\end{abstract}

\section{Introduction}
\label{sec:introduction}
Microscopes are widely used in biological and medical research, allowing the study of organic and inorganic substances at the minuscule scale. The observed microscopy images, however, suffer from the following two inherent distortions: a blur of detail caused by the resolution limit of a microscope, and a background noise introduced by the imperfections of the imaging system as a whole and by the image-recording sensor in particular. Both of these traits not only distort the perception of the detail in the image but also influence the quantitative analysis of its content~\cite{Conchello2005FluoM,Monvel2001ImageRF}.

Mathematically, the image formation process can be described by the operation of convolution, where the underlying image is convolved with the point spread function of the microscope~\cite{Evangelista2005FromCT,Sheppard1980ImageFI}. Two fluorescence microscopy imaging systems -- widefield and confocal -- are very popular among biologists. In the case of widefield microscopy, the entire specimen is exposed to a uniform light source. Here, the intensity of the illuminating light is high and the noise statistics of the recorded image can be approximated by the Gaussian distribution~\cite{wu2010microscope}. On the contrary, the confocal microscopes rely on point-by-point imaging of the specimen thanks to an aperture (pin-hole) installed in the microscope's optical system to block the out-of-focus signal. This improves the resolution but limits the numerical aperture of the microscope, effectively reducing the number of photons captured by the imaging detector. A more appropriate approximation for the noise in such low-photon images is the Poisson distribution~\cite{wu2010microscope}.

Restoration of microscopy images is an ill-posed inverse problem where a unique solution does not exist~\cite{Evangelista2005FromCT}. One, therefore, needs to constrain the space of solutions in order to obtain a statistically or a physically meaningful one. A popular approach for doing that follows the variational formulation of the problem, where the restored image is obtained as the minimizer of an objective function~\cite{2013IntroductionTV}. This function comprises two terms: the data fidelity term that measures the proximity between the obtained measurements and the solution, and the regularization term that integrates prior information about the expected solution. There are several methods focused on the development of an effective regularization scheme for the image restoration, e.g., the Hyper-Laplacian priors~\cite{Krishnan2009FastID}, the non-local means~\cite{Buades2005ANA}, the shrinkage fields~\cite{Schmidt2014ShrinkageFF}, and others.

With the advent of deep learning, many inverse problems have been successfully approached with Fully Convolutional Neural Networks (FCNNs)~\cite{Mildenhall2017BurstDW,Tao2017DetailRevealingDV} which can learn a mapping between the measured image and its expected reconstruction. Furthermore, a series of works have incorporated deep learning for the regularization purposes in a wide range of image restoration problems \cite{Zhang2016LearningFC,Lefkimmiatis_2017_CVPR,Lefkimmiatis_2018_CVPR,Zhang2017LearningDC,Kruse2017LearningTP,Kokkinos2018DeepID}. Such a regularization paradigm, in turn, has allowed to surpass the performance of the one-shot methods relying on FCNN; however, none of these techniques have been tested on microscopy related deblurring problems.

In this work, we present a joint denoising and deblurring framework for microscopy images, which comprises a collection of methods that leverage the advantages of both the classical schemes for optimization, and the deep learning approaches for regularization. We develop an extensive set of techniques for handling the image prior information, by using both shallow and deep learning for parametrization. The proposed framework entails the following steps:

\begin{itemize}
\vspace{-.2cm}
    \item[-] First, a regularizer is formed as a group of learnable kernels and is deployed to every image identically.
    \item[-] Second, an intuitive extension is further examined with the group of kernels being predicted \emph{per image} using a compact FCNN as a Kernel Prediction Network (KPN).
    \item[-] The same kernel-predicting approach is then probed in \emph{per pixel} evaluation, with the KPN predicting the appropriate regularizer for each spatial location in the image. 
    \item[-] The last step consists of approximating the entire regularization function with a neural network, which is then employed in an iterative manner.
    \vspace{-.2cm}
\end{itemize}

The developed methods outperform solutions based either on optimization schemes alone or the solutions based merely on deep learning techniques. Exploiting both the classical optimization and the deep learning methodologies, the proposed approaches are intrinsically ready for fine-tuning the trade-off between the computational efficiency and the accuracy of image reconstruction.

\section{Related Work}
\vspace{-.1cm}
Deconvolution in the presence of noise is generally considered a challenging task, and it has attracted significant attention by the research community. 
The simplest deconvolution approach consists in estimation of the maximum likelihood under the assumption of the statistical model of the observed data. The main drawback of this approach is the amplification of the measurement noise. One common way to avoid this hurdle is to add the regularization term to penalize the values of the solution.

One of the most well-known classical methods for image deblurring and denoising, the Wiener-Kolmogorov filter, or Wiener filter~\cite{Wiener1949TheEI,Tintner1980LinearLE}, is derived as a maximum likelihood estimator under the assumption of Gaussian noise. It uses the Tikhonov regularization functional to incorporate prior knowledge about the expected solution.  Wiener filter also retrieves the minimum mean squared error estimate and it has established itself as a fast deconvolution algorithm~\cite{MedicalDeblurHussein2016,Boyat2014ImageDU,Zhang2016AnES} due to the closed-form solution in the Fourier domain. More recently, the Wiener filter became a part of methods such as PURE-LET~\cite{Li2018PURELETID} and SURE-LET~\cite{Zhang2016AnES}, performing deconvolution with a piece-wise thresholding under wavelet coefficients regularization. 

Another well-known method for image deblurring is the Richardson-Lucy algorithm, which is derived as a maximum-likelihood estimator under the assumption of Poisson noise \cite{Richardson1972BayesianBasedIM,Lucy1974AnIT}. Richardson-Lucy algorithm is a common method for tackling problems of image deblurring and denoising in microscopy and it is mostly used with Tikhonov regularization and total-variation regularization \cite{Dey2006RichardsonLucyAW}.

In order to drive the solution of the optimization problem towards a subset of physically plausible image reconstructions, many researchers have resorted to hand-crafted regularization schemes. For example, Krishnan \etal in~\cite{Krishnan2009FastID} propose to use hyper-Laplacian penalty functions, while in~\cite{Beek2010ImageDA} a method using non-local regularization constraint is proposed for image deblurring and denoising.  

With the advent of deep learning, nearly all image restoration methods were revisited from a learnable perspective with great success. Being widely used in many research areas, including the biomedical field, a plethora of neural network approaches have been proposed for denoising~\cite{Mildenhall2017BurstDW}, demosaicking~\cite{Kokkinos_2019}, and super-resolution~\cite{Tao2017DetailRevealingDV}. Deep learning methods are also applied in image deblurring, e.g., in~\cite{Eigen2013RestoringAI}, where a CNN is used for restoring images corrupted with various visual artifacts, and in~\cite{Chowdhury2017}, where a parametric CNN model was used to enhance a shape-based artifact elimination. Xu \etal in~\cite{Xu2014DeepCN} proposed a CNN that learns the deconvolution operation for natural images in a supervised manner. 
A step into combining traditional optimization schemes with deep learning is made in~\cite{Schmidt2014ShrinkageFF} by using iterative FFT-based deconvolution with learnable regularization filters, weights and shrinkage functions. A similar approach was introduced by~\cite{Zhang2016LearningFC}, where the authors proposed to learn horizontal and vertical image gradient filters. Zhang \etal in~\cite{Zhang2017LearningDC} developed a trainable denoiser prior that is then integrated into a model-based optimization method. Finally, in~\cite{Bigdeli2017DeepMP} the authors propose to learn a prior that represents a Gaussian-smoothed version of the natural image distribution.
 
\vspace{-.2cm}
\section{Problem Formulation}
\vspace{-.1cm}
\subsection{Image Formation in Microscopy}

The image formation process can be  described by the observation model
\begin{equation}\label{eq:model}
\mathbf{y} = \mathbf{K x} + \mathbf{n},
\end{equation}
where $\mathbf{y} \in \mathbb{R}^{N}$ corresponds to the observed image, $\mathbf{K} \in \mathbb{R}^{N \times N}$ is the matrix corresponding to the point spread function (PSF), $\mathbf{x} \in \mathbb{R}^{N}$ is the underlying image that we aim to restore and $\mathbf{n} \sim \mathcal{N}(0,\,\sigma^2)$ denotes noise, which is assumed to follow i.i.d Gaussian distribution. While $\mathbf{x}$ and $\mathbf{y}$ are two dimensional images, for the sake of mathematical derivations, we assume that they have been raster scanned using a lexicographical order, and they correspond to vectors of $\textit{N}$ dimensions.

\vspace{-.1cm}
\subsection{Regularization}

Deconvolution being an image restoration task is an ill-posed inverse problem, which implies that a unique solution does not exist. In general, such problems can be addressed following a variational approach whose solution $\hat{\mathbf{x}}$ can be obtained by minimizing an objective function of the form
\begin{equation}\label{eq:estimate}
\hat{\mathbf{x}} = \arg\!\min_{\m x} \underbrace{\frac{1}{2}||\mathbf{y} - \mathbf{Kx}||^{2}_{2} + \lambda r({\mathbf{x}})}_{\mathbf{J(x)}}.
\end{equation}
Here the first term corresponds to the data fidelity term, which measures the proximity of the solution to the observation, while the second one corresponds to the regularizer that models any prior knowledge one might have about the ground-truth image. The parameter $\lambda$ is a trade-off coefficient that determines the contribution of the regularizer into the estimation of the solution. A list of popular regularizers contains Tikhonov \cite{Tikhonov:1963} and TV functionals \cite{rudin1992nonlinear} which have been widely used in a plethora of image restoration tasks, including deconvolution problems \cite{Dey2006RichardsonLucyAW}. However, in this work we attempt to incorporate prior information learned directly from available training data in a supervised manner with the inclusion of deep learning strategies.

\begin{figure*}[t!] 
\label{fig:method_viz}
\vspace{-.3cm}
\captionsetup{format=plain, font=small, labelfont=bf}
\captionsetup[subfigure]{position=bottom}
  \centering
  \subfloat[WF-K \label{fig:viz_WF-K}]{\includegraphics[width=.46\textwidth]{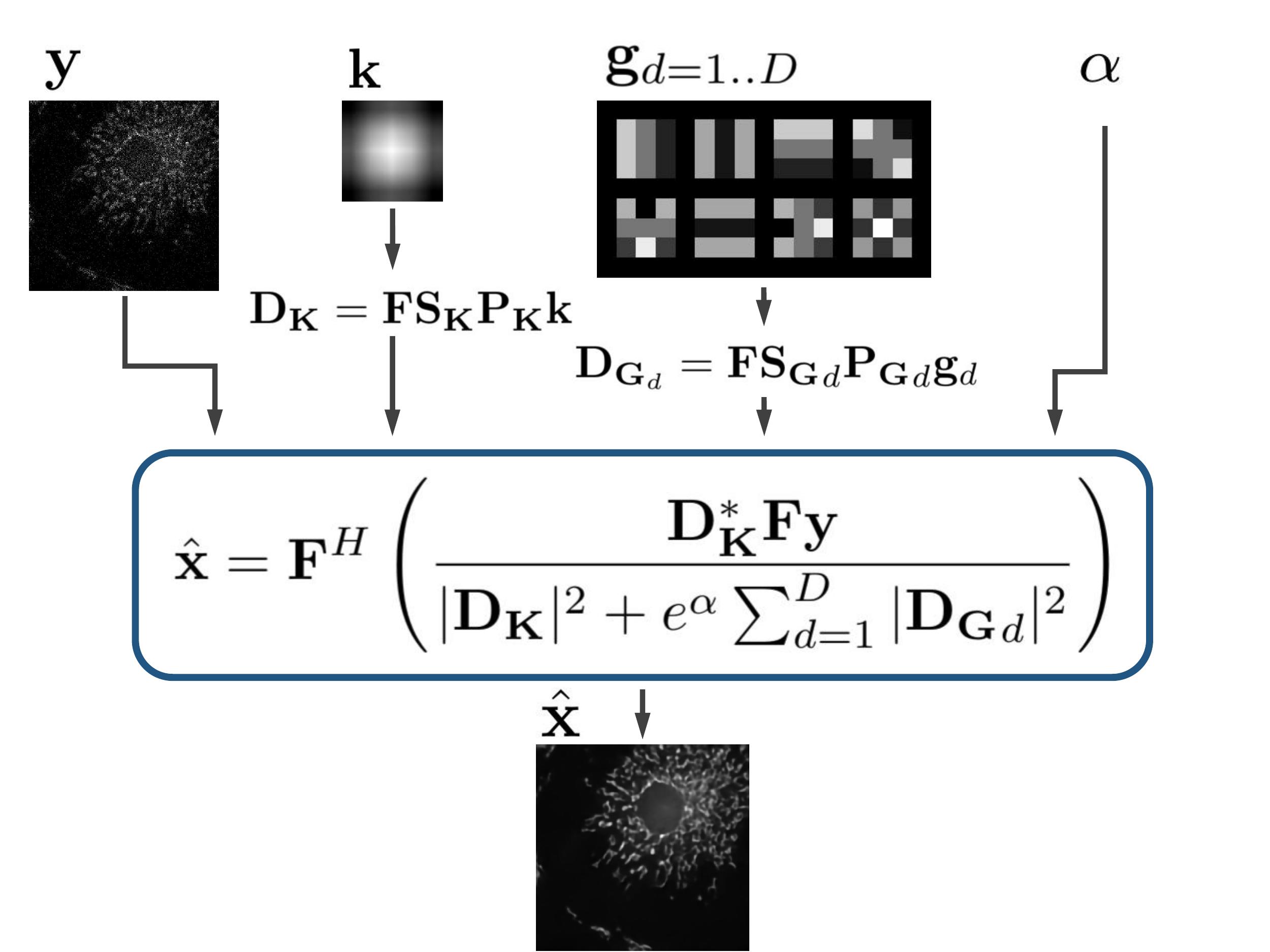}} 
  \rulesep
  \subfloat[WF-KPN \label{fig:viz_WF-KPN}]{\includegraphics[width=.46\textwidth]{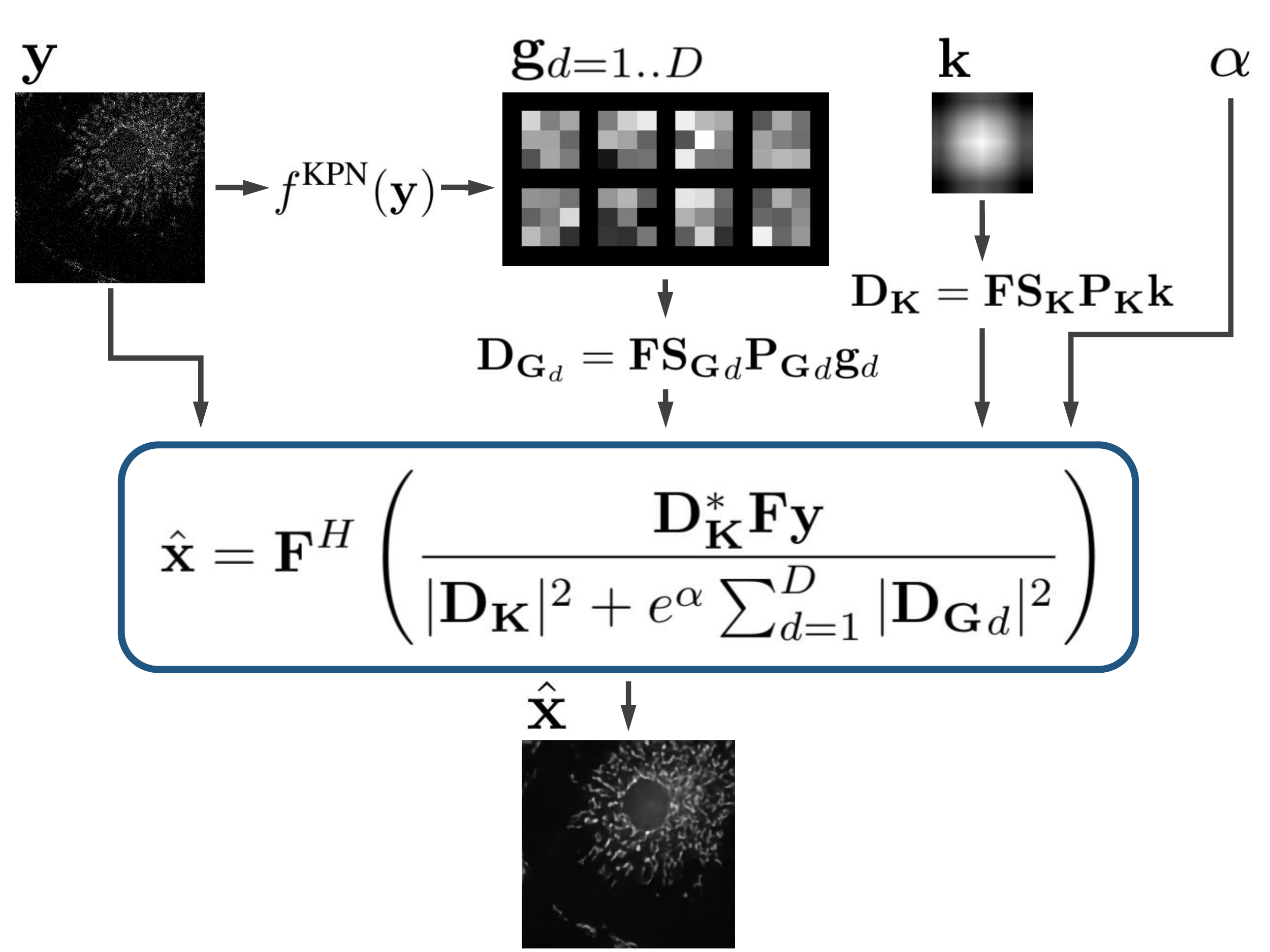}} \\
  \subfloat[WF-KPN-SA \label{fig:viz_WF-KPN-SA}]{\includegraphics[width=.46\textwidth]{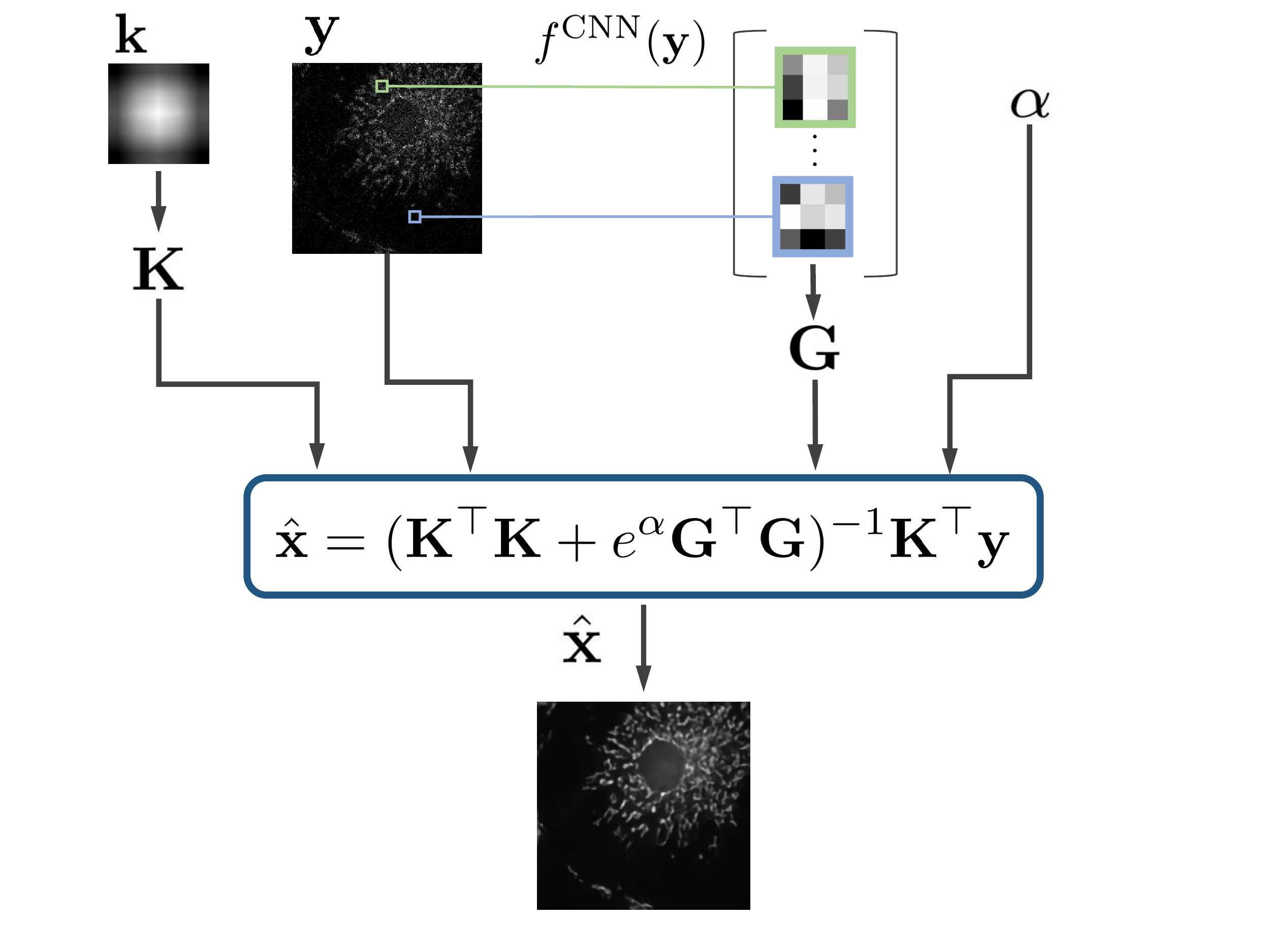}}
  \rulesep
  \subfloat[WF-UNet \label{fig:viz_WF-UNet}]{\includegraphics[width=.46\textwidth]{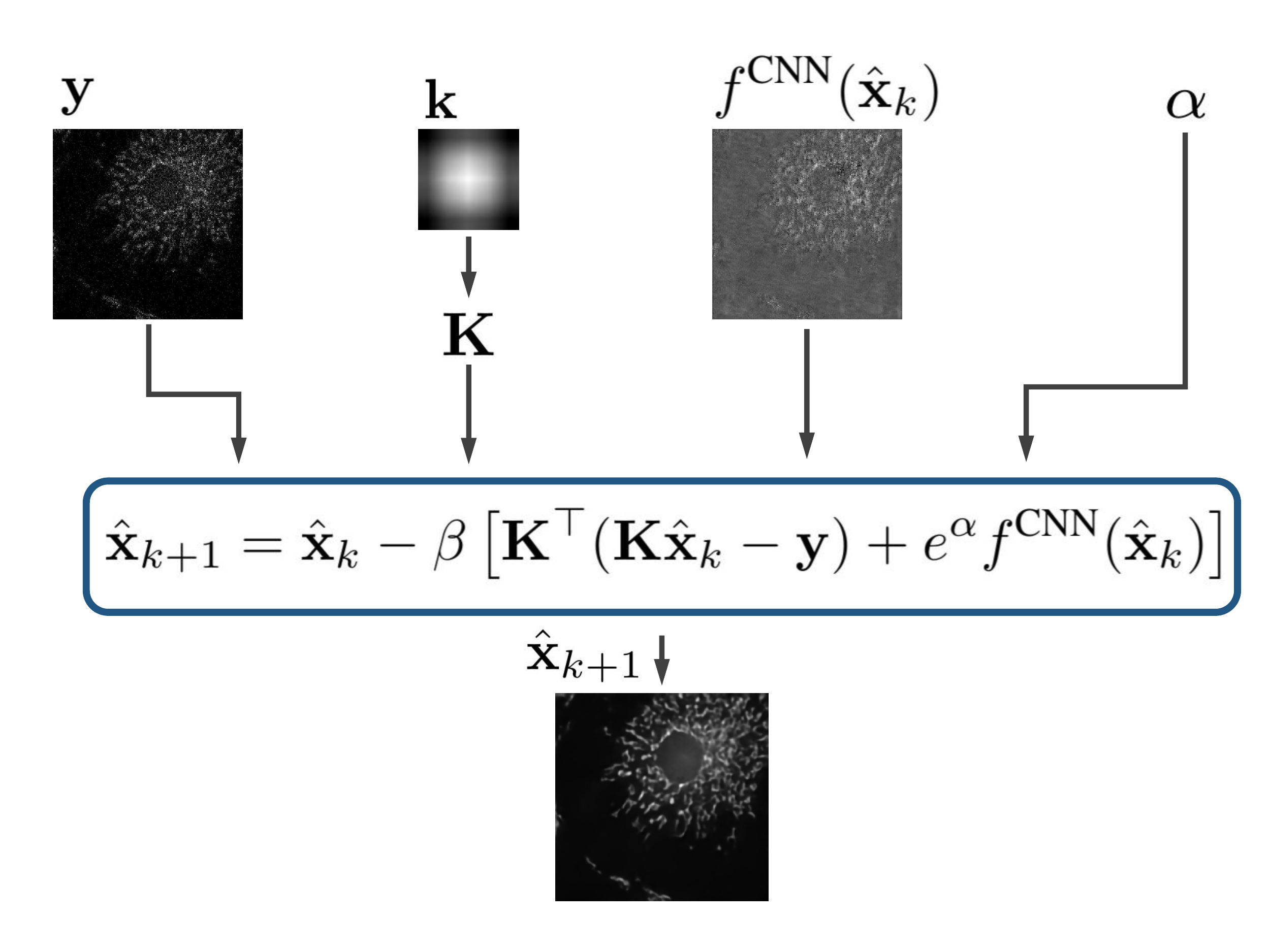}}
  \vspace{.1cm}
  \caption{Overview of the four proposed prior parametrization models. In Fig.~\ref{fig:viz_WF-K} the one-shot Wiener filter with learnable kernels is depicted, while in~\ref{fig:viz_WF-KPN} the same filter with predictable kernels is presented. In~\ref{fig:viz_WF-KPN-SA} the scheme incorporating prediction of per-pixel regularization kernels into Wiener filter is shown. The proposed iterative scheme which combines Wiener filtering with the approximation of regularization with a CNN is visualized in~\ref{fig:viz_WF-UNet}} 
\end{figure*}

\section{Proposed Approach}
\label{sec:all_models}
\subsection{Learnable Regularization Kernels}

First, we start from the observation that in most of the modern methods for image restoration the regularization term is formed as
\begin{equation}\label{eq:regularizer}
r({\mathbf{x}}) = \sum_{d=1}^{D}\rho_{d}(\mathbf{g}_{d} * \mathbf{x}),
\vspace{-.1cm}
\end{equation}
where $\mathbf{g}_{d}$ are typically linear filters and $\rho_{d}(\cdot)$ is a set of penalty functions, acting on the filters outputs \cite{Krishnan2009FastID}. In this work, we explicitly set the penalty function to be the squared $\ell_2$ norm, leading to a Tikhonov regularizer~\cite{Tikhonov:1963} 
\vspace{-.1cm}
\begin{equation}\label{eq:prior_kernels}
r(\mathbf{x}) = \sum_{d=1}^{D}||\mathbf{g}_{d} * \mathbf{x}||^{2}_{2} = \sum_{d=1}^{D}||\mathbf{G}_{d}\mathbf{x}||^{2}_{2}, 
\vspace{-.1cm}
\end{equation}
where $\mathbf{g}_{d}$ are learnable convolution kernels and $\mathbf{G}_{d} \in \mathbb{R}^{N \times N}$ are convolution matrices, corresponding to these kernels. This specific choice of the penalty function allows us to obtain the solution of \eqref{eq:estimate} in the closed-form
\begin{equation}\label{eq:wiener}
\hat{\mathbf{x}} = (\mathbf{K}^{\top}\mathbf{K} + \lambda\sum_{d=1}^{D}\mathbf{G}_{d}^{\top}\mathbf{G}_{d})^{-1}\mathbf{K}^{\top}\mathbf{y},
\end{equation}
which corresponds to the Wiener-Kolmogorov filter.
Here $\mathbf{K}^{\top}$ and $\mathbf{G}_{d}^{\top}$ are the adjoint matrices of $\mathbf{K}$ and $\mathbf{G}_{d}$, respectively. 
In order to obtain the solution of Eq.~\eqref{eq:wiener} one has first to perform the inversion of a huge matrix which can be very slow in practice or even intractable. However, the supremacy of the Wiener filter lies in its FFT-based inference, which renders it a fast and efficient method and allows to restore the underlying signal at a low computational complexity. Specifically, under the assumption of periodic image boundary conditions, the degradation matrix $\mathbf{K}$ and convolution matrix $\mathbf{G}_{d}$ are circulant real matrices and thus, they can be diagonalized in the Fourier domain as
\begin{equation}
\mathbf{K} = \mathbf{F}^{H}\mathbf{D}_{\mathbf{K}}\mathbf{F},\ \mathbf{G}_{d} = \mathbf{F}^{H}\mathbf{D}_{\mathbf{G}_{d}}\mathbf{F}
\vspace{-.45cm}
\label{eq:circ}
\end{equation}

\begin{equation}\label{eq:diag}
\mathbf{D}_{\mathbf{K}} = \mathbf{FS_{K}P_{K}k},\ 
\mathbf{D}_{\mathbf{G}_{d}} = \mathbf{FS_{G}}_{d}\mathbf{P_{G}}_{d}\mathbf{g}_{d}.
\end{equation}
Here $\mathbf{F} \in \mathbb{C}^{N \times N}$ is the Fourier (DFT) matrix,  $\mathbf{F}^{H} \in \mathbb{C}^{N \times N}$ is its inverse, $\mathbf{D}_{\mathbf{K}}$, $\mathbf{D}_{\mathbf{G}_{d}} \in \mathbb{C}^{N \times N}$ are diagonal matrices, $\mathbf{S_{K}}, \mathbf{S_{G}}_{d} \in \mathbb{R}^{N \times N}$ are the corresponding circulant shift operators, $\mathbf{P_{K}} \in \mathbb{R}^{N \times M}, \mathbf{P_{G}}_{d} \in \mathbb{R}^{N \times Ld}$ are the corresponding zero-padding operators, $\mathbf{k} \in \mathbb{R}^{M}$ is the blurring kernel and $\mathbf{g}_{d} \in \mathbb{R}^{Ld}$ is a regularization convolution kernel. We also consider the trade-off coefficient $\lambda$ to be equal to $e^{\alpha}$ in all experiments to ensure the positivity of the regularization weight. Based on the above, the one-shot solution of Eq.~\eqref{eq:wiener} is
\begin{equation}\label{eq:wiener_fourier}
\hat{\mathbf{x}} = \mathbf{F}^{H} \left ( \frac{\mathbf{D^{*}_{K}}\mathbf{F}\mathbf{y}}{|\mathbf{D_{K}}|^{2} + e^{\alpha}\sum_{d=1}^{D}|\mathbf{D_{G}}_{d}|^{2}} \right ),
\end{equation}
where $\mathbf{D^{*}_{K}}$ is the Hermitian transpose of $\mathbf{D_{K}}$ and the division is applied in an element-wise fashion. The method is depicted in Fig.~\ref{fig:viz_WF-K} and hereafter we refer to it as WF-K. To obtain the solution in this form, we firstly employ a group of $D$ learnable kernels $\mathbf{g}_{d}$ of size $K\times K$  which are initialized using a two-dimensional discrete cosine transform (DCT) frequency basis; a common choice in image processing since it has shown the ability to extract useful prior image information. We also consider $\alpha$ to be a learnable parameter in all proposed regularization schemes in order for the trade-off coefficient value to be tuned alongside the learnable kernels during the training process.

\subsection{Prediction of Regularization Kernels} \label{sec:wf_kpn}
A critical drawback of the standard Wiener method lies in the way the group of kernels is formulated. In detail, the learnable group is global which means that all kernels are applied on every image identically without any dependency on the underlying content. To remedy this situation, we propose the content-driven WF-KPN method that predicts per image the group of kernels that need to be used for regularization. In this framework, the solution of a Wiener filter yields the same form as presented in Eq.~\eqref{eq:wiener_fourier}, but unlike the previous approach, the convolution kernels $\mathbf{g}_{d}$, that form the diagonal matrix $\mathbf{D_{G}}_{d}$ in Eq.~\eqref{eq:diag}, are now predicted from a Kernel Prediction Network (KPN)~\cite{Mildenhall2017BurstDW} as shown in Fig.~\ref{fig:viz_WF-KPN}.

For the Kernel Prediction Network we select a compact customized UNet architecture with nearly 470k parameters that receives the distorted input $\mathbf{y}$ and produces an output with $K^{2}D$ channels and the same spatial resolution as the input. Unlike the traditional UNet, we perform global average pooling in the spatial dimension on the output of the last layer, which is then reshaped into a stack of $D \times K \times K$ regularization kernels. Here $D$ is the number of kernels and $K \times K$ is the support size. In all our experiments, we set $D = 8$ and $K = 3$. We also use instance normalization \cite{Ulyanov2016InstanceNT} after each convolutional layer to normalize weights for each image independently. In this way, the KPN predicts $D$ content dependent regularization kernels of size $K \times K$ for each image, that then under the assumption of periodic boundary conditions form the diagonal matrix $\mathbf{D}_{\mathbf{G}_{d}}$ in Eq.~\eqref{eq:diag}. After that, the resulting diagonal matrices $\mathbf{D}_{\mathbf{G}_{d}}$ are incorporated into the solution of the Wiener filter according to Eq.~\eqref{eq:wiener_fourier}.

The proposed models WF-K and WF-KPN involve learning of the trade-off coefficient $\alpha$ and regularization kernels $\mathbf{g}_{d}$ in a supervised manner via means of back-propagation of a loss function. While the loss function we are minimizing during training is real-valued, the solution in Eq.~\eqref{eq:wiener_fourier} involves complex quantities and therefore we cannot perform back-propagation by solely relying on layers currently available in existing deep-learning libraries. For this reason we have implemented our own customized layers which depend on the analytical derivations of the gradients of the solution $\mathbf{\hat{x}}$ w.r.t. the trainable parameters. All analytical derivations of the necessary gradients are provided in the supplementary material.

\subsection{Prediction of Spatially Adaptive Regularization Kernels}
While the prediction of global regularization kernels per image provides the appealing property of content adaptation, we further extend WF-KPN to be both spatially and content adaptive. This is achieved by predicting for each image a different regularization kernel per-pixel and hence the method is dubbed as WF-KPN-SA. For this extension, we modify the UNet of Section~\ref{sec:wf_kpn} to predict a kernel per spatial location of an image. Unlike WF-KPN, the output of the network has $K^{2}$ channels and the same size as the input. We empirically found that normalization of the network hinders the performance and therefore it was removed in all WF-KPN-SA related experiments. Furthermore, the output of the network is reshaped into spatially adaptive regularization kernels of size $K \times K$ for each pixel of an input image. These kernels are then unfolded into a matrix $\mathbf{G}$ which now does not correspond to a circulant matrix, and, thus, the FFT-based inference of the Wiener-Kolmogorov filter is not feasible. The solution of the restoration problem has now the form of
\vspace{-.1cm}
\begin{equation}
\hat{\mathbf{x}} = (\mathbf{K}^{\top}\mathbf{K} + e^{\alpha}\mathbf{G}^{\top}\mathbf{G})^{-1}\mathbf{K}^{\top}\mathbf{y}
\label{eq:wiener_for_cg}
\vspace{-.15cm}
\end{equation}
and the calculation of $\hat{\mathbf{x}}$ is done using the conjugate gradient algorithm~\cite{Shewchuk_1994}. To the best of our knowledge, WF-KPN-SA is the first method that predicts a spatially varying regularizer per image for inverse imaging related  tasks. 

\subsection{Prediction of the Gradient of the Regularizer}

Finally, with the desire to fully exploit deep learning capabilities, we employ a CNN for parametrizing a prior that would be specific for each image and, thus, completely content adaptive. This way we do not make assumptions about the form of the regularization term as we did above, and thus, the solution of Eq.~\eqref{eq:estimate} cannot be derived anymore in a closed-form expression. One common way to solve Eq.~\eqref{eq:estimate} in this case is to apply a gradient descent~\cite{Bertsekas1999NonlinearPS} optimization algorithm to find a minimizer of the objective function $\mathbf{J(x)}$,
\begin{equation}\label{eq:gradient_descent}
    \hat{\mathbf{x}}_{k+1} = \hat{\mathbf{x}}_{k} - \beta \nabla \mathbf{J(\mathbf{x})}.
\end{equation}
Here $\hat{\mathbf{x}}_{k+1}$ is the solution of Eq.~\eqref{eq:estimate} that is updated after each iteration $k$ of the gradient descent scheme and $\beta$ is the learning rate defining the speed of the algorithm. Introducing the objective function $\mathbf{J(x)}$, defined according to Eq.~\eqref{eq:estimate}, into Eq.~\eqref{eq:gradient_descent}, we obtain the solution of the gradient descent scheme,
\vspace{-.1cm}
\begin{equation}
    \hat{\mathbf{x}}_{k+1} = \hat{\mathbf{x}}_{k} - \beta \left[\mathbf{K}^{\top}(\mathbf{K}\hat{\mathbf{x}}_{k} - \mathbf{y}) + e^{\alpha} f^{\text{CNN}}(\hat{\mathbf{x}}_{k}) \right].
    \vspace{-.1cm}
\label{eq:wf_unet}
\end{equation}
Here we parametrize the gradient of the regularizer with the CNN, $\nabla\mathbf{r}(\hat{\mathbf{x}}_{k}) = f^{\text{CNN}}(\hat{\mathbf{x}}_{k})$. For the prediction of the gradient of a regularizer we employ the UNet architecture that was defined in Section~\ref{sec:wf_kpn}. Unlike WF-KPN and WF-KPN-SA, we do not modify the original UNet architecture, and in this case the network receives a distorted image $\mathbf{y}$ as an input and maps it to an output with the same resolution and number of channels as the input. We also use instance normalization after each convolutional layer to normalize weights for each image independently. With a slight abuse of notation, the method is hereafter referred as WF-UNet and it is depicted in~Fig.~\ref{fig:viz_WF-UNet}. The number of gradient descent iterations for WF-UNet is set to be equal to 10. We also consider the step size $\beta$ to be a learnable parameter in order for the gradient descent speed to be tuned during the training process.

We stress that the proposed methods, WF-KPN, WF-KPN-SA and WF-UNet, share a UNet with almost identical architecture in order to have approximately the same number of trainable parameters that allows for fair comparison between the models. Furthermore, the selected network architecture is relatively compact, which in return allows the development of methods capable of deblurring an image in milliseconds. In detail, the architecture of the implemented UNet consists of a contracting path and an expansive path. Each step of the contracting path involves application of 2 convolutional layers of $3 \times 3$ kernels each followed by a $2 \times 2$ max pooling operation for downsampling. The number of feature channels starts from 12 and increases at each downsampling except the last one by a factor of 2 reaching a value of 96. Each step of the expanding path of the network consists of an upsampling, concatenation with the feature map from the contracting path and 2 convolutional layers of $3 \times 3$ kernels each. Respectively, the number of channels decreases at each upsampling except the last one by a factor of 2. The rectified linear unit is applied after each convolutional layer except the final convolutional layer that produces the output of the network.

\begin{figure*}[t]
  \centering
  \subfloat{\includegraphics[width=1.\textwidth]{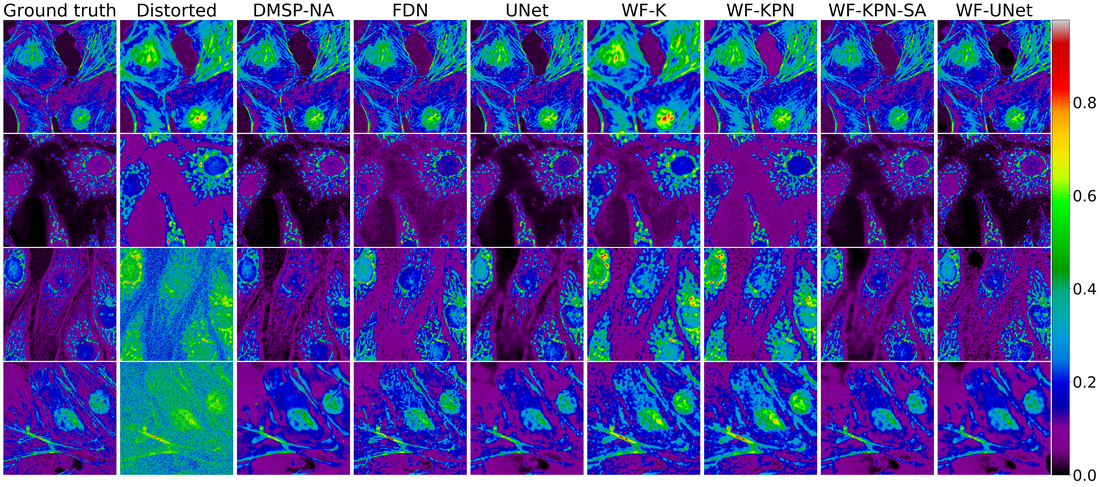}}
\captionsetup{format=plain, font=small, labelfont=bf}
\caption{Restoration of microscopy images degraded by PSF and Gaussian noise with standard deviation equal to 0.005, 0.01, 0.05, 0.1, respectively, from top to bottom. All images are originally grayscale but a different colormap is used to better highlight the differences among the various reconstructions} 
  \label{fig:deblur_gaussian}
\end{figure*}

\vspace{-.1cm}
\section{Network Training}

\subsection{Dataset}
\label{sec:dataset}

Since there is a lack of a deconvolution dataset, comprising of microscopy images perturbed by Gaussian noise, we create training pairs of ground-truth and distorted with blur and noise images using the Fluorescence Microscopy Denoising (FMD) dataset~\cite{Zhang_2019_CVPR} and a dataset that is used in cell segmentation of microscopy images~\cite{AlKofahi2018ADL}. To create a large set of reference images, we take the ground-truth images from both datasets and crop each image into patches of size $256 \times 256$. Consequently, all patches that contain no information about cells are discarded by comparing the mean value of a patch with the mean value of the original image. Note that all ground-truth images as well as resulting distorted images are grayscale.

To produce blurred versions of the ground-truth images we create 35 different 2D PSFs of sizes $7 \times 7$, $9 \times 9$, $11 \times 11$ and $13 \times 13$ by altering the index of refraction of the media, the numerical aperture, the pixel size and the excitation wavelength using the ImageJ plugins Diffraction PSF 3D~\cite{DP3D} and PSF Generator~\cite{PSFG} with the Richards \& Wolf optical model. In detail, 25 PSFs are reserved for training, 5 for validation and the rest for testing purposes. Notably, all PSFs $\mathbf{k}$ are normalized such that $\sum_{i}k_{i} = 1$, which is a realistic assumption in optics \cite{Dey2006RichardsonLucyAW}. 

To assess the performance of our algorithms on a wide range of Gaussian noise levels, all images are scaled in the range of [0, 1], then blurred with a PSF and finally perturbed with i.i.d. Gaussian noise with standard deviation from the set (0.001, 0.005, 0.01, 0.05, 0.1). 

The resulting dataset, which contains 1405 pairs of ground-truth and distorted images, is split into 975 training, 200 validation and 230 testing samples. During training, a ground-truth sample is convolved with a randomly chosen blur kernel from the 25 training PSFs, and subsequently Gaussian noise with a randomly selected standard deviation from the set presented above is added to the blurred image.

\subsection{Training Details}

All proposed methods are trained in an end-to-end fashion to minimize the $\ell_1$ loss function between the output and the ground-truth image as well as the gradients of the aforementioned entities
\begin{equation}\label{eq:loss}
\mathcal{L} = ||\hat{\mathbf{x}} - \mathbf{x}||_{1} + ||\nabla \hat{\mathbf{x}} - \nabla \mathbf{x}||_{1}.
\end{equation}
It has been reported that training with the $\ell_1$ loss function yields images with sharper edges \cite{Zhao2017LossFF}. Also, by incorporating a gradient based loss alongside the pixel-wise loss, we obtain models that are capable of reconstructing the sharp details of the underlying images. All proposed methods are optimized using Adam~\cite{Kingma2014AdamAM} with a learning rate $10^{-3}$. For all methods, except the WF-KPN-SA, we use batch size equal to 25 and train the models for 300 epochs. For the WF-KPN-SA network, due to memory-related limitations, we set the batch size to be equal to 3. 

\subsection{Evaluation}
\label{sec:Evaluation}

\begin{table*}[t]
\centering
\captionsetup{format=plain, font=small, labelfont=bf}
\caption{PSNR and SSIM comparisons on Gaussian image deblurring for five noise levels. In all tables, we highlight the best result in bold and underline the second best result\label{tab:comparison_gaussian}}
\vspace{.3cm}
\small
\resizebox{12cm}{!} {
\begin{tabular}{lllllllllll}
\specialrule{.1em}{.05em}{.05em} 
& \multicolumn{10}{c}{STD} \\
\specialrule{.1em}{.05em}{.05em}  
& \multicolumn{2}{c}{0.001} & \multicolumn{2}{c}{0.005} & \multicolumn{2}{c}{0.01} & \multicolumn{2}{c}{0.05} & \multicolumn{2}{c}{0.1}\\
& \multicolumn{1}{c}{PSNR} & \multicolumn{1}{c}{SSIM} & \multicolumn{1}{c}{PSNR} & \multicolumn{1}{c}{SSIM} & \multicolumn{1}{c}{PSNR} & \multicolumn{1}{c}{SSIM} & \multicolumn{1}{c}{PSNR} & \multicolumn{1}{c}{SSIM} & \multicolumn{1}{c}{PSNR} & \multicolumn{1}{c}{SSIM} \\ \hline
\multicolumn{1}{l|}{Input} & \multicolumn{1}{l}{36.23} & \multicolumn{1}{l|}{.8955} & \multicolumn{1}{l}{35.37} & \multicolumn{1}{l|}{.8791} & \multicolumn{1}{l}{33.93} & \multicolumn{1}{l|}{.8339} & \multicolumn{1}{l}{26.03} & \multicolumn{1}{l|}{.3858} & \multicolumn{1}{l}{21.14} & .1718 \\
\multicolumn{1}{l|}{IRCNN~\cite{Zhang2017LearningDC}} & 33.33 & \multicolumn{1}{l|}{.8604} & 36.88 & \multicolumn{1}{l|}{.8972} & 36.80 & \multicolumn{1}{l|}{.9013} & 32.44 & \multicolumn{1}{l|}{.7932} & 28.95 & .6835 \\ 
\multicolumn{1}{l|}{FDN~\cite{Kruse2017LearningTP}} & \underline{40.31} & \multicolumn{1}{l|}{\textbf{.9424}} & 38.61 & \multicolumn{1}{l|}{.9239} & 37.33 & \multicolumn{1}{l|}{.9086} & {33.50} & \multicolumn{1}{l|}{{{.8406}}} & {30.49} & {.7550} \\ 
\multicolumn{1}{l|}{DMSP-NA~\cite{Bigdeli2017DeepMP}} & \textbf{40.44} & \multicolumn{1}{l|}{.9402} & \textbf{39.16} & \multicolumn{1}{l|}{\textbf{{.9290}}} & \underline{37.73} & \multicolumn{1}{l|}{\underline{{.9123}}} & \underline{34.31} & \multicolumn{1}{l|}{\underline{{.8589}}} & {32.19} & .8312 \\
\multicolumn{1}{l|}{DMSP-NB~\cite{Bigdeli2017DeepMP}} & 40.27 & \multicolumn{1}{l|}{\underline{.9411}} & {37.23} & \multicolumn{1}{l|}{{.9027}} & {36.59} & \multicolumn{1}{l|}{{.8930}} & 34.29 & \multicolumn{1}{l|}{{.8582}} & {32.20} & \underline{.8313} \\
\hline
\multicolumn{1}{l|}{UNet} & 37.73 & \multicolumn{1}{l|}{.9177} & 37.27 & \multicolumn{1}{l|}{.9136} & 36.64 & \multicolumn{1}{l|}{.9064} & 34.04 & \multicolumn{1}{l|}{.8556} & \underline{32.40} & .8187 \\
\multicolumn{1}{l|}{WF-K} & 35.66 & \multicolumn{1}{l|}{.8849} & 35.61 & \multicolumn{1}{l|}{.8834} & 35.45 & \multicolumn{1}{l|}{.8787} & 32.74 & \multicolumn{1}{l|}{.7950} & 29.27 & .6835 \\
\multicolumn{1}{l|}{WF-KPN} & 38.72 & \multicolumn{1}{l|}{.9253} & 37.98 & \multicolumn{1}{l|}{.9176} & 36.80 & \multicolumn{1}{l|}{.9028} & 32.33 & \multicolumn{1}{l|}{.8022} & 29.20 & .7259 \\
\multicolumn{1}{l|}{WF-KPN-SA} & {39.86} & \multicolumn{1}{l|}{{{.9390}}} & \underline{38.76} & \multicolumn{1}{l|}{\underline{.9275}} & \textbf{37.81} & \multicolumn{1}{l|}{\textbf{.9157}} & \textbf{34.58} & \multicolumn{1}{l|}{\textbf{.8688}} & \textbf{32.60} & \textbf{.8363} \\
\multicolumn{1}{l|}{WF-UNet} & 38.08 & \multicolumn{1}{l|}{.9102} & 37.55 & \multicolumn{1}{l|}{.9053} & 36.77 & \multicolumn{1}{l|}{.8966} & 33.89 & \multicolumn{1}{l|}{.8442} & 32.01 & .8096 \\
\specialrule{.1em}{.05em}{.05em} 
\end{tabular}
}
\vspace{-.5cm}
\end{table*}

To evaluate the performance of all algorithms on each noise level, we use 230 ground-truth images that are convolved with the 5 PSFs reserved for testing purposes. Note that the held-out PSFs were not seen during training in order to explore the generalization ability of the methods to unknown blurring kernels. Finally, the noisy observation of a blurry image is produced by adding Gaussian noise with each standard deviation from the set (0.001, 0.005, 0.01, 0.05, 0.1). In detail, we get 1150 test samples in total, combined into 5 test sets of different noise levels which allows the thorough comparison of the performance of our algorithms on each noise level separately.

We compare our proposed algorithms with three state-of-the-art algorithms for deblurring images distorted by Gaussian noise: IRCNN \cite{Zhang2017LearningDC}, FDN \cite{Kruse2017LearningTP} and DMSP~\cite{Bigdeli2017DeepMP}. The latter was considered in two implementations -- noise-blind, or noise-adaptive (NA), and noise-aware, or non-blind (NB).
We also provide a CNN baseline which is a UNet with the same architecture and  number of parameters as the one used for the parametrization of the regularization gradient. This way we get a valid comparison and we are able to investigate if the combination of a standard optimization scheme and a deep learning approach will help to achieve good restoration results and surpass those obtained by employing solely a deep learning based algorithm. For assessing the methods performance we use the standard peak signal-to-noise ratio (PSNR) and the structural similarity index (SSIM~\cite{Wang2004ImageQA}) metrics. Comparisons of all the methods runtime was conducted on a computer with an Intel Core i7-8750H CPU and a NVIDIA GeForce GTX 1080Ti GPU.

\vspace{-.2cm}
\section{Results}

We evaluate all methods on the developed test set across all different noise ranges, as described in Section~\ref{sec:Evaluation}. The results presented in Table~\ref{tab:comparison_gaussian} show that our proposed spatially adaptive method WF-KPN-SA surpasses in performance all competing state-of-the-art methods except in the very low noise cases. The performance over the deep learning based competing methods  FDN~\cite{Kruse2017LearningTP}, IRCNN~\cite{Zhang2017LearningDC} and DMSP~\cite{Bigdeli2017DeepMP} ranges from nearly 0.1 in the low noise regime to 0.4 dB for high noise. Only in the case of very low noise the previous state-of-the-art methods provide better reconstruction quality than WF-KPN-SA. However, we stress that, as well as DMSP-NA, our methods are noise-blind, while IRCNN, FDN and DMSP-NB are noise-aware, with the oracle noise standard deviation being explicitly provided as an input.

Furthermore, we find that Wiener Filter (WF-K) with the learnable regularization filters yields satisfactory reconstruction quality and perform only slightly worse than several deep learning methods for a fraction of its computational complexity. A detailed benchmark of computational time for all methods presented in our work can be found in Table~\ref{tab:comp_cost}.  
As presented in Fig.~\ref{fig:deblur_gaussian}, WF-K and WF-KPN methods tend to miss sharp details in the restored images, while WF-KPN-SA and WF-UNet allow the restoration of fine image details.

\begin{table}[t]
\centering
\captionsetup{format=plain, font=small, labelfont=bf}
\caption{Runtime of several deconvolution algorithms for processing an image with the spatial dimensions of $256 \times 256$. All times were calculated using the publicly available implementations and the reported benchmarks are the average of 10 runs}
\vspace{.3cm}
\small
\begin{tabular}{lcclcc}
\specialrule{.1em}{.05em}{.05em} 
 & CPU, ms & GPU, ms &  & CPU, ms & GPU, ms\\ \hline
\multicolumn{1}{l|}{GILAM \cite{Chen2014RegularizedGI}} & \multicolumn{1}{c|}{4155.9} & \multicolumn{1}{c|}{--} & \multicolumn{1}{l|}{UNet} & \multicolumn{1}{c|}{32.4} & \textbf{2.8} \\
\multicolumn{1}{l|}{HSPIRAL \cite{Lefkimmiatis2013PoissonIR}} & \multicolumn{1}{c|}{5828.8} & \multicolumn{1}{c|}{--} & \multicolumn{1}{l|}{WF-K} & \multicolumn{1}{c|}{\textbf{5.6}} & 4.6 \\
\multicolumn{1}{l|}{PURE-LET \cite{Li2018PURELETID}} & \multicolumn{1}{c|}{211.3} & \multicolumn{1}{c|}{--}  & \multicolumn{1}{l|}{WF-KPN} & \multicolumn{1}{c|}{45.4} & 7.7\\
\multicolumn{1}{l|}{FDN \cite{Kruse2017LearningTP}} & \multicolumn{1}{c|}{34542.1} & \multicolumn{1}{c|}{1095.0} & \multicolumn{1}{l|}{WF-KPN-SA} & \multicolumn{1}{c|}{968.2} & 341.4  \\
\multicolumn{1}{l|}{DMSP \cite{Bigdeli2017DeepMP}} & \multicolumn{1}{c|}{79656.0} & \multicolumn{1}{c|}{19707.7} & \multicolumn{1}{l|}{WF-UNet} & \multicolumn{1}{c|}{354.1} & 39.8 \\
\multicolumn{1}{l|}{IRCNN \cite{Zhang2017LearningDC}} & \multicolumn{1}{c|}{7890.7} & \multicolumn{1}{c|}{--} & \multicolumn{1}{l|}{} & \multicolumn{1}{c|}{} &  \\
\specialrule{.1em}{.05em}{.05em} 
\end{tabular}
\label{tab:comp_cost}
\vspace{-.3cm}
\end{table}

\vspace{-.1cm}
\section{Poisson Image Deblurring}

As an extension, all developed methods are applied in the case of confocal microscopy, where as it was mentioned in Section~\ref{sec:introduction} the intensity of the light illuminating the sample is very small and thus the noise statistics obey the Poisson distribution. One notable property of the Poisson distribution is that the mean and variance of the random variable are not independent, i.e. $\text{mean}(\m y) = \text{var}(\m y)$. As such, the image formation model is formulated as $\mathbf{y} = \mathcal{P}(\mathbf{K x})$ where  $\mathcal{P}$ denotes the Poisson noise distorting the image. One successful way to perform deconvolution and denoising in the presence of Poisson noise is to first apply a variance stabilizing transformation (VST)~\cite{Mkitalo2013OptimalIO} which transforms a variable from the Poisson distribution into one from the Gaussian. This allows to borrow the ample apparatus of the well-studied methods derived for Gaussian statistics. This approach has been successfully applied in the literature \cite{Mkitalo2013OptimalIO,Foi2008PracticalPN,Zhang_2019_CVPR,Lu20013DDW} to solve the Poissonian restoration problem using Gaussian denoising algorithms. 

In our work we explore a similar approach and make use of the widely known VST, the Anscombe transform \cite{Anscombe1948THETO}, which is applied on the distorted observation
\vspace{-.2cm}
\begin{equation}\label{eq:anscombe}
\mathbf{y}\rightarrow 2\sqrt{\mathbf{y} + \frac{3}{8}}.
\end{equation}
The Anscombe transform aims to stabilize the data variance to be approximately unity. After the transformation, the data can be viewed as a signal-independent Gaussian process with unit variance, and therefore the Wiener filter, which is derived from Gaussian statistics, can be directly applied. Once the deconvolved solution, denoted as $\hat{\mathbf{x}}$, is obtained, an inverse Anscombe transform is applied to return the data to its original domain. Applying the simple algebraic inverse usually results in a biased estimate of the output. To mitigate the bias in the case of photon-limited imaging the exact unbiased inverse transformation should be used, whose closed-form approximation~\cite{Mkitalo2011ACA} is 
\begin{equation}\label{eq:exact_unbiased}
\hat{\mathbf{x}} \rightarrow \Big(\frac{\hat{\mathbf{x}}}{2}\Big)^{2} - \frac{1}{8} + \frac{1}{4}\sqrt{\frac{3}{2}}\hat{\mathbf{x}}^{-1} -\frac{11}{8}\hat{\mathbf{x}}^{-2}+\frac{5}{8}\sqrt{\frac{3}{2}}\hat{\mathbf{x}}^{-3}.
\end{equation}

\subsubsection{Data Preparation and Results}
\begin{table*}[t]
\centering
\captionsetup{format=plain, font=small, labelfont=bf}
\caption{PSNR and SSIM comparisons on Poisson image deblurring for six different noise levels. Methods that were not able to produce meaningful results due to numerical instability issues are marked with N/A\label{tab:comparison_Poisson}} 
\vspace{.3cm}
\small
\resizebox{12.5cm}{!} {
\begin{tabular}{lllllllllllll}
\specialrule{.1em}{.05em}{.05em} 
 & \multicolumn{12}{c}{PEAK} \\
 \specialrule{.1em}{.05em}{.05em}  
 & \multicolumn{2}{c}{1} & \multicolumn{2}{c}{2} & \multicolumn{2}{c}{5} & \multicolumn{2}{c}{10} & \multicolumn{2}{c}{25} & \multicolumn{2}{c}{50} \\
 & \multicolumn{1}{c}{PSNR} & \multicolumn{1}{c}{SSIM} & \multicolumn{1}{c}{PSNR} & \multicolumn{1}{c}{SSIM} & \multicolumn{1}{c}{PSNR} & \multicolumn{1}{c}{SSIM} & \multicolumn{1}{c}{PSNR} & \multicolumn{1}{c}{SSIM} & \multicolumn{1}{c}{PSNR} & \multicolumn{1}{c}{SSIM} & \multicolumn{1}{c}{PSNR} & \multicolumn{1}{c}{SSIM} \\ \hline
\multicolumn{1}{l|}{Input} & \multicolumn{1}{l}{10.93} & \multicolumn{1}{l|}{.0981} & \multicolumn{1}{l}{13.22} & \multicolumn{1}{l|}{.1104} & \multicolumn{1}{l}{16.67} & \multicolumn{1}{l|}{.1639} & \multicolumn{1}{l}{19.45} & \multicolumn{1}{l|}{.2425} & \multicolumn{1}{l}{23.21} & \multicolumn{1}{l|}{.3930} & \multicolumn{1}{l}{26.00} & \multicolumn{1}{l}{.5247} \\
\multicolumn{1}{l|}{GILAM~\cite{Chen2014RegularizedGI}} & 24.57 & \multicolumn{1}{l|}{.4977} & 25.66 & \multicolumn{1}{l|}{.6350} & 26.64 & \multicolumn{1}{l|}{.5451} & 28.24 & \multicolumn{1}{l|}{.6304} & 29.92 & \multicolumn{1}{l|}{.6975} & 30.97 & .7463 \\
\multicolumn{1}{l|}{HSPIRAL~\cite{Lefkimmiatis2013PoissonIR}} & 23.18 & \multicolumn{1}{l|}{.4101} & 26.54 & \multicolumn{1}{l|}{.5809} & 30.18 & \multicolumn{1}{l|}{.7576} & 31.94 & \multicolumn{1}{l|}{.8229} & 33.39 & \multicolumn{1}{l|}{.8578} & 34.38 & .8729 \\
\multicolumn{1}{l|}{PURE-LET~\cite{Li2018PURELETID}} & 26.18 & \multicolumn{1}{l|}{.7318} & 26.44 & \multicolumn{1}{l|}{.7494} & 26.77 & \multicolumn{1}{l|}{.7612} & 27.67 & \multicolumn{1}{l|}{.7799} & 28.77 & \multicolumn{1}{l|}{.0.8065} & 28.77 & .8114 \\
\multicolumn{1}{l|}{FDN~\cite{Kruse2017LearningTP}} & N/A & \multicolumn{1}{l|}{N/A} & N/A & \multicolumn{1}{l|}{N/A} & N/A & \multicolumn{1}{l|}{N/A} & N/A & \multicolumn{1}{l|}{N/A} & N/A & \multicolumn{1}{l|}{N/A} & 31.16 & .8176 \\ 
\multicolumn{1}{l|}{DMSP-NA~\cite{Bigdeli2017DeepMP}} & 16.07 & \multicolumn{1}{l|}{0.4243} & 18.86 & \multicolumn{1}{l|}{.5648} & 23.80 & \multicolumn{1}{l|}{.7126} & 27.76 & \multicolumn{1}{l|}{.7880} & 31.74 & \multicolumn{1}{l|}{.8490} & 33.65 & .8770 \\ 
\multicolumn{1}{l|}{DMSP-NB~\cite{Bigdeli2017DeepMP}} & 10.86 & \multicolumn{1}{l|}{.2079} & 12.97 & \multicolumn{1}{l|}{.2761} & 20.49 & \multicolumn{1}{l|}{.5691} & 26.12 & \multicolumn{1}{l|}{.7356} & 30.25 & \multicolumn{1}{l|}{.8136} & 33.01 & .8633 \\ 
\multicolumn{1}{l|}{IRCNN~\cite{Zhang2017LearningDC}} & 6.81 & \multicolumn{1}{l|}{.0902} & 11.67 & \multicolumn{1}{l|}{.1809} & 17.37 & \multicolumn{1}{l|}{.3695} & 22.14 & \multicolumn{1}{l|}{.5323} & 29.46 & \multicolumn{1}{l|}{.8038} & 33.10 & .8660 \\ \hline
\multicolumn{1}{l|}{UNet} & 28.50 & \multicolumn{1}{l|}{ .7916} & 29.79 & \multicolumn{1}{l|}{.8143} & 31.47 & \multicolumn{1}{l|}{.8428} & 32.72 & \multicolumn{1}{l|}{.8633} & 34.20 & \multicolumn{1}{l|}{.8871} & 35.17 & .9008 \\
\multicolumn{1}{l|}{WF-K} & 25.81 & \multicolumn{1}{l|}{.5821} & 27.85 & \multicolumn{1}{l|}{.6947} & 29.94 & \multicolumn{1}{l|}{.7888} & 31.13 & \multicolumn{1}{l|}{.8288} & 32.06 & \multicolumn{1}{l|}{.8557} & 32.43 & .8654 \\
\multicolumn{1}{l|}{WF-KPN} & 27.09 & \multicolumn{1}{l|}{.7503} & 28.19 & \multicolumn{1}{l|}{.7811} & 29.89 & \multicolumn{1}{l|}{.8177} & 31.34 & \multicolumn{1}{l|}{.8460} & 33.22 & \multicolumn{1}{l|}{.8770} & 34.50 & .8945 \\
\multicolumn{1}{l|}{WF-KPN-SA} & {\ul 28.80} & \multicolumn{1}{l|}{{\ul .7949}} & {\ul 30.12} & \multicolumn{1}{l|}{{\ul.8185}} & {\ul 31.72} & \multicolumn{1}{l|}{{\ul .8468}} & {\ul 32.92} & \multicolumn{1}{l|}{{\ul .8672}} & {\ul 34.48} & \multicolumn{1}{l|}{{\ul .8909}} & {\ul 35.60} & {\ul.9056} \\
\multicolumn{1}{l|}{WF-UNet} & \textbf{29.04} & \multicolumn{1}{l|}{\textbf{.8005}} & \textbf{30.27} & \multicolumn{1}{l|}{\textbf{.8220}} & \textbf{31.83} & \multicolumn{1}{l|}{\textbf{.8492}} & \textbf{33.06} & \multicolumn{1}{l|}{\textbf{.8694}} & \textbf{34.59} & \multicolumn{1}{l|}{ \textbf{.8928}} & \textbf{35.67} & {\textbf{.9069}} \\
\specialrule{.1em}{.05em}{.05em} 
\end{tabular}
}
\vspace{-.3cm}
\end{table*}

\begin{figure*}[t]
 \centering
 \captionsetup{format=plain, font=small, labelfont=bf}
  \subfloat{\includegraphics[width=1.\textwidth]{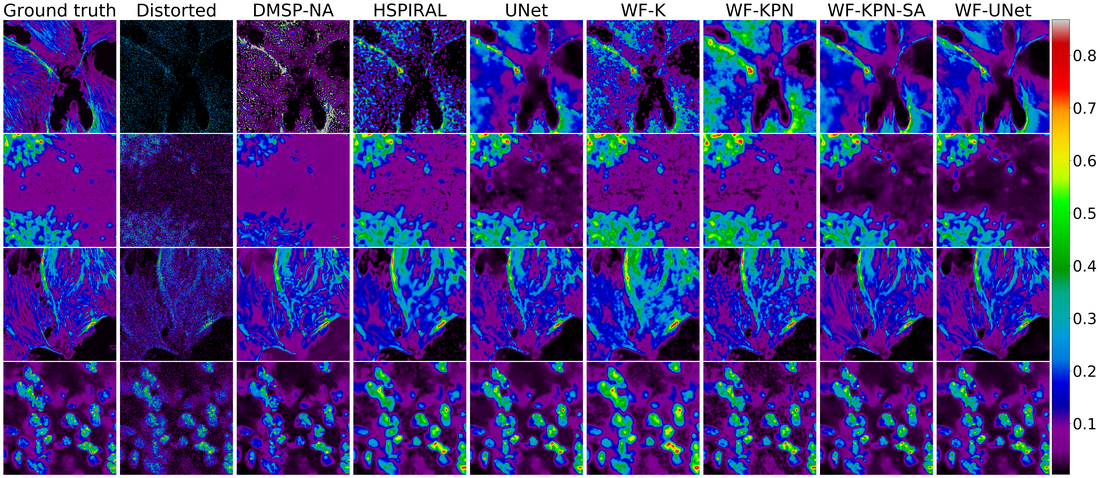}}
  \caption{Restoration of microscopy images scaled to have peak intensities equal to 1, 5, 10, 25, respectively from top to bottom and degraded by PSF and Poisson noise. All images are originally grayscale, but a different colormap is used to better highlight the differences among the various reconstructions \label{fig:deblur_Poisson}}
\end{figure*}

To assess the performance of our algorithms on the task of Poisson image deblurring, we retrain our models using the VST and the exact unbiased inverse transformation included into the pipeline. The Anscombe transform is applied on the input image $\mathbf{y}$ as described in~Eq.~\eqref{eq:anscombe} before being fed into any of the proposed models of Section~\ref{sec:all_models}. Accordingly, the restored signal $\hat{\mathbf{x}}$ from any of the proposed models is transformed back using the exact unbiased version of the inverse transformation, presented in Eq.~\eqref{eq:exact_unbiased}, in order to recover the final solution. For this task we use the same dataset as the one used for the evaluation of the Gaussian deblurring. We create 35 2D PSFs of size $5 \times 5$ to simulate confocal microscope pinhole with the ImageJ Diffraction PSF 3D plugin \cite{PSFG}. We again use 25 PSFs for training purposes, 5 for validation and 5 for test. To simulate various SNR values of Poisson noise, we use the prior art approach \cite{Chen2014RegularizedGI,Lefkimmiatis2013PoissonIR} and scale the ground-truth images to have a maximum intensity of (1, 2, 5, 10, 25, 50). Poisson noise is signal dependent with local $\text{SNR} =\sqrt{y_{i}}$, where $y_{i}$ denotes the underlying image intensity at position \textit{i}, therefore by increasing the maximum intensity of an image, the amount of noise decreases and vice versa. Furthermore, it is long known fact that images with large mean value and perturbed with Poisson noise follow approximately a normal, or Gaussian distribution, and therefore methods based on Gaussian statistics might work equally well. Scaling the ground-truth images to have various ranges of maximum intensities aims to cover a wide gamut of noise levels, including the ones that belong to approximately a Gaussian distribution. 

During training, a ground-truth sample is rescaled to a randomly chosen maximum intensity from the range mentioned above. Then the sample is convolved with a randomly chosen blur kernel from the 25 training PSFs, and after that the noisy observation is produced from the blurred image. We follow the same strategy as in Section \ref{sec:Evaluation} for producing blurred and noisy observations for the test set.

We compare the proposed algorithms with the non-blind deblurring algorithms, developed for Poisson distribution: GILAM \cite{Chen2014RegularizedGI}, HSPIRAL \cite{Lefkimmiatis2013PoissonIR} and PURE-LET \cite{Li2018PURELETID}. We also include the aforementioned state-of-the-art methods for Gaussian image deblurring into the comparison by applying the VST to the distorted data and the exact unbiased transformation to the result of the restorations.
Table~\ref{tab:comparison_Poisson} and Fig.~\ref{fig:deblur_Poisson} clearly show that WF-UNet outperforms the other approaches at most of the noise levels, which is especially important on the low intensity peak values where the Poisson noise is stronger. WF-KPN-SA shows comparable to WF-UNet performance being only marginaly inferior. Overall, WF-UNet and WF-KPN-SA show good quantitative improvement from nearly 2.6dB to 5.8dB over the previous state-of-the-art methods on the low intensity peaks and from nearly 1.2dB to 7dB on the high intensity peaks. Moreover, Fig.~\ref{fig:deblur_Poisson} demonstrates that, despite the good quantitative improvement, WF-K and WF-KPN struggle to reconstruct the fine details in the images, while the reconstructions generated by WF-KPN-SA and WF-UNet methods are very accurate.

\vspace{-.2cm}
\section{Conclusion}

In this work, we proposed a series of methods based on the Wiener-Kolmogorov filtering technique for dealing with the problem of Gaussian and Poisson image deblurring. We introduced three novel ways to parametrize the image priors, including a new approach of regularization with kernel predictions obtained by a neural network. Our extensive experimentation line showcased that our proposed framework based on the prediction of a regularizer achieves superior quality of image reconstruction and surpasses the solutions that rely either on deep learning or on optimization schemes alone. Finally, several of our proposed algorithms demonstrate low computational complexity, without sacrificing the accuracy of image restoration. Being fast and accurate, the proposed framework paves the way towards real-time microscopy image restoration.

\bibliographystyle{splncs04}
\bibliography{egbib}

\clearpage

\begin{center}
      {\textbf{\Large Microscopy Image Restoration with Deep Wiener-Kolmogorov Filters -- Supplementary Material}}\\
       \vspace{2ex}
    \end{center}

\setcounter{equation}{0}
\setcounter{figure}{0}
\setcounter{section}{0}
\setcounter{table}{0}
\setcounter{page}{1}

\section{Derivation of expressions for backward pass for WF-K and WF-KPN}

Under the assumption of periodic image boundary conditions, the degradation matrix $\mathbf{K}$ and the convolution matrix $\mathbf{G}_{d}$ are circulant and, therefore, they can be diagonalized in the Fourier domain. Hence, the solution of the Wiener filter for the proposed models WF-K and WF-KPN can be expressed in a closed form as
\begin{equation}\label{eq:wiener_fourier}
\hat{\mathbf{x}} = \mathbf{F}^{H} \left ( \frac{\mathbf{D^{*}_{K}}\mathbf{F}\mathbf{y}}{|\mathbf{D_{K}}|^{2} + e^{\alpha}\sum_{d=1}^{D}|\mathbf{D_{G}}_{d}|^{2}} \right ),
\end{equation}
where the division of the vector in the numerator by the diagonal matrix in the denominator is applied in an element-by-element fashion.

The regularization kernels $\mathbf{g}_{d}$ and the power of the trade-off $\alpha$ are learned during the training process using back-propagation. While the solution to Eq.~\ref{eq:wiener_fourier} comprises complex quantities, the loss function used for the network training is real-valued, allowing the parameter update to be performed using back-propagation. For this reason, we have implemented our customized layers which depend on the analytical derivations of the gradients of the solution $\hat{\mathbf{x}}$ w.r.t. the trainable parameters. 

Below, we present the derivation of the expressions of the gradients for the backward pass. Unless explicitly stated otherwise, we assume the use of denominator-layout notation in the calculations.

\subsection{Gradient w.r.t. $\alpha$}

We start the derivation of the closed-form expressions for back-propagation with a calculation of the gradient of the Wiener filter solution $\hat{\mathbf{x}}$ w.r.t. the parameter $\alpha$. Let us denote the solution in Eq.~\eqref{eq:wiener_fourier} as a function defined by the parameters,
\begin{equation}
  f(\alpha, \mathbf{y}, \mathbf{g}_{d})  = \mathbf{F}^{H} \left ( \frac{\mathbf{D^{*}_{K}}\mathbf{F}\mathbf{y}}{|\mathbf{D_{K}}|^{2} + e^{\alpha}\sum_{d=1}^{D}|\mathbf{D_{G}}_{d}|^{2}} \right ).
\label{eq:2}
\end{equation}
Note that similarly to Eq.~\eqref{eq:wiener_fourier}, the division operation in the Eq.~\eqref{eq:2} is meant to describe the element-wise division of the vector in the numerator by the corresponding diagonal element of the diagonal matrix in the denominator.

For simplicity and compactness of the calculations we denote the numerator and the denominator in Eq.~\eqref{eq:2} as
\begin{align}
&\mathbf{D^{*}_{K}}\mathbf{F}\mathbf{y} = \mathbf{z}, \label{eq:numer} \\
&|\mathbf{D_{K}}|^{2} + e^{\alpha}\sum_{d=1}^{D}|\mathbf{D_{G}}_{d}|^{2} = \mathbf{\Omega}.
\label{eq:denom}
\end{align}
With the notations defined in Eqs.~\eqref{eq:numer} and \eqref{eq:denom}, the function in Eq.~\eqref{eq:2} yields the form of
\begin{equation}
f(\alpha, \mathbf{y}, \mathbf{g}_{d}) = \mathbf{F}^{H}\mathbf{\Omega}^{-1}\mathbf{z}.
\label{eq:3}
\end{equation}
Note that $\mathbf{\Omega}^{-1}$ is a diagonal matrix. Assuming the numerator-layout notation, the gradient of $f(\alpha, \mathbf{y}, \mathbf{g}_{d})$ w.r.t. the parameter $\alpha$ can be calculated as
\begin{align}
\begin{split}
\frac{\partial f(\alpha, \mathbf{y}, \mathbf{g}_{d})}{\partial \alpha} 
&= \frac{\partial \mathbf{F}^{H}\mathbf{\Omega}^{-1}\mathbf{z}}{\partial\alpha} = - \mathbf{F}^{H}\mathbf{\Omega}^{-1}\frac{\partial \mathbf{\Omega}}{\partial \alpha}\mathbf{\Omega}^{-1}\mathbf{z} = \\
&= - \mathbf{F}^{H}\mathbf{\Omega}^{-1}e^{\alpha}\sum_{d=1}^{D}|\mathbf{D_{G}}_{d}|^{2}\mathbf{\Omega}^{-1}\mathbf{z} = - e^{\alpha}\mathbf{F}^{H}\mathbf{\Omega}^{-2}\sum_{d=1}^{D}|\mathbf{D_{G}}_{d}|^{2}\mathbf{D}^{*}_{\mathbf{K}}\mathbf{Fy}.
\end{split}
\label{eq:4}
\end{align}

\subsection{Gradient w.r.t. $\mathbf{g}_{d}$}

Next, we derive the expression for the gradient of $f(\alpha, \mathbf{y}, \mathbf{g}_{d})$ w.r.t. the regularization kernel $\mathbf{g}_{d}$. To do that, we rewrite Eq.~\eqref{eq:2} as $f(\alpha, \mathbf{y}, \mathbf{g}_{d}) = \mathbf{F}^{H}\mathbf{h}(\mathbf{g})$, where
\begin{equation}
\mathbf{h}(\mathbf{g}) = \frac{\mathbf{\text{\boldmath$\lambda$}}^{*}_{\mathbf{K}} \odot \mathbf{Fy}}{ |\mathbf{\text{\boldmath$\lambda$}}_{\mathbf{K}}|^{2} + e^{\alpha}\sum_{d=1}^{D}|\mathbf{T}_{d}\mathbf{g}_{d}|^{2}}.
\label{eq:5}
\end{equation}
This way, we can express the gradient of $f(\alpha, \mathbf{y}, \mathbf{g}_{d})$ w.r.t. $\mathbf{g}_{d}$ as
\begin{equation}
\frac{\partial f(\alpha, \mathbf{y}, \mathbf{g}_{d})}{\partial \mathbf{g}_{d}} = \frac{\partial \mathbf{F}^{H} \mathbf{h}(\mathbf{g})}{\partial \mathbf{g}_{d}} = \frac{\partial \mathbf{h}(\mathbf{g})}{\partial \mathbf{g}_{d}}\mathbf{F}^{*}.
\label{eq:11}
\end{equation} We emphasize that $\mathbf{D_{K}}$ and $\mathbf{D}_{\mathbf{G}_{d}}$ are diagonal matrices and we use the following notation:
\begin{align}
&\mathbf{\text{\boldmath$\lambda$}}_{\mathbf{K}} = \textit{vec}(\mathbf{D_{K}}) \nonumber \\
&\mathbf{\text{\boldmath$\lambda$}}_{\mathbf{G}_{d
}} = \textit{vec}(\mathbf{D}_{\mathbf{G}_{d
}}) = \mathbf{T}_{d}\mathbf{g}_{d} \label{eq:notations} \\
&\mathbf{T}_{d} = \mathbf{FS_{G}}_{d}\mathbf{P_{G}}_{d} \in \mathbb{C}^{N \times Ld}.  \nonumber
\end{align}
Here $\textit{vec}(\mathbf{D})$ is meant to define the vector which lies on the main diagonal of the diagonal matrix $\mathbf{D}$. Note that the operation $\odot$ corresponds to element-wise multiplication, and the division in Eq.~\eqref{eq:5} applies element-wisely. Here $\mathbf{S_{G}}_{d} \in \mathbb{R}^{N \times N}$ is a circulant shift operator and $\mathbf{P_{G}}_{d} \in \mathbb{R}^{N \times Ld}$ is a zero-padding operator.

Taking into consideration Eq.~\eqref{eq:5} and the notations in Eq.~\eqref{eq:notations}, $\mathbf{h}(\mathbf{g})$ can be rewritten as
\begin{equation}
\mathbf{h}(\mathbf{g}) =
\begin{bmatrix}
\frac{\mathbf{M}_{1}\mathbf{\text{\boldmath$\lambda$}}^{*}_{\mathbf{K}} \odot \mathbf{Fy}}{|\mathbf{M}_{1}\mathbf{\text{\boldmath$\lambda$}}_{\mathbf{K}}|^{2} + e^{\alpha}\sum_{d=1}^{D}|\mathbf{M}_{1}\mathbf{T}_{d}\mathbf{g}_{d}|^{2}}\\ 
...\\
\frac{\mathbf{M}_{N}\mathbf{\text{\boldmath$\lambda$}}^{*}_{\mathbf{K}} \odot \mathbf{Fy}}{|\mathbf{M}_{N}\mathbf{\text{\boldmath$\lambda$}}_{\mathbf{K}}|^{2} + e^{\alpha}\sum_{d=1}^{D}|\mathbf{M}_{N}\mathbf{T}_{d}\mathbf{g}_{d}|^{2}}
\end{bmatrix},
\label{eq:6}
\end{equation}
where $\mathbf{M}_{i} \in \mathbb{R}^{1 \times N}$ is a vector every element of which except $i$ is equal to zero, and the $i$-th element is equal to 1. Therefore, the $i$-th element of the vector $\mathbf{h(g)}$ is
\begin{equation}
\mathbf{h}_{i}(\mathbf{g}) =
\frac{\mathbf{M}_{i}\mathbf{\mathbf{\text{\boldmath$\lambda$}}}^{*}_{\mathbf{K}} \odot \mathbf{Fy}}{|\mathbf{M}_{i}\mathbf{\mathbf{\text{\boldmath$\lambda$}}}_{\mathbf{K}}|^{2} + e^{\alpha}\sum_{d=1}^{D}|\mathbf{M}_{i}\mathbf{T}_{d}\mathbf{g}_{d}|^{2}} = \frac{\mathbf{a}_{i}}{\mathbf{b}_{i} + e^{\alpha}\mathbf{u}_{i}(\mathbf{g})}. 
\label{eq:7}
\end{equation}
Here we denote 
\begin{align}
    &\mathbf{a}_{i} = \mathbf{M}_{i}\mathbf{\text{\boldmath$\lambda$}}^{*}_{\mathbf{K}} \odot\mathbf{Fy}, \nonumber \\
    &\mathbf{b}_{i} = |\mathbf{M}_{i}\mathbf{\text{\boldmath$\lambda$}}_{\mathbf{K}}|^{2} \label{eq:notations2} \\
    &\mathbf{u}_{i} = \sum_{d=1}^{D}|\mathbf{M}_{i}\mathbf{T}_{d}\mathbf{g}_{d}|^{2} = \sum_{d=1}^{D}\mathbf{g}^{\top}_{d}\mathbf{T}^{*\top}_{d}\mathbf{M}^{\top}_{i}\mathbf{M}_{i}\mathbf{T}_{d}\mathbf{g}_{d} \nonumber.
\end{align} 
To calculate the gradient of $\mathbf{h}_{i}(\mathbf{g})$ w.r.t. the $j$-th regularization kernel $\mathbf{g}_{j}$ one must perform indirect differentiation, starting with acquiring the gradient of $\mathbf{u}_{i}(\mathbf{g})$ w.r.t. $\mathbf{g}_{j}$, 
\begin{align}
\begin{split}
\frac{\partial \mathbf{u}_{i}(\mathbf{g})}{\partial \mathbf{g}_{j}} &=
\sum_{d=1}^{D}\left (\frac{\partial{\mathbf{g}_{d}}}{\partial{\mathbf{g}_{j}}}\mathbf{T}^{*\top}_{d}\mathbf{M}^{\top}_{i}\mathbf{M}_{i}\mathbf{T}_{d}\mathbf{g}_{d} + \frac{\partial{\mathbf{g}_{d}}}{\partial{\mathbf{g}_{j}}}\mathbf{T}^{\top}_{d}\mathbf{M}^{\top}_{i}\mathbf{M}_{i}\mathbf{T}^{*}_{d}\mathbf{g}_{d}\right ) =
\\
&= (\mathbf{T}^{*\top}_{j}\mathbf{M}^{\top}_{i}\mathbf{M}_{i}\mathbf{T}_{j} + \mathbf{T}^{\top}_{j}\mathbf{M}^{\top}_{i}\mathbf{M}_{i}\mathbf{T}^{*}_{j})\mathbf{g}_{j}.
\label{eq:8}
\end{split}
\end{align}
Incorporating the result obtained in Eq.~\eqref{eq:8} into $\frac{\partial \mathbf{h}_{i}(\mathbf{g})}{\partial \mathbf{g}_{j}}$, we derive
\begin{align}
\begin{split}
\frac{\partial \mathbf{h}_{i}(\mathbf{g})}{\partial \mathbf{g}_{j}} &= \mathbf{a}_{i} \frac{\partial (\mathbf{b}_{i} + e^{\alpha}\mathbf{u}_{i}(\mathbf{g}))^{-1}}{\partial \mathbf{g}_{j}} = - \mathbf{a}_{i} \frac{e^{\alpha}}{(\mathbf{b}_{i} + e^{\alpha}\mathbf{u}_{i}(\mathbf{g}))^{2}}\frac{\partial \mathbf{u}_{i}(\mathbf{g})}{\partial \mathbf{g}_{j}} = \\
&= - e^{\alpha} \frac{\mathbf{M}_{i}\mathbf{\text{\boldmath$\lambda$}}^{*}_{\mathbf{K}} \odot \mathbf{Fy}}{(|\mathbf{M}_{i}\mathbf{\text{\boldmath$\lambda$}}_{\mathbf{K}}|^{2} + e^{\alpha}\sum_{d=1}^{D}|\mathbf{M}_{i}\mathbf{T}_{d}\mathbf{g}_{d}|^{2})^{2}}\cdot (\mathbf{R}{ij} + \overline{\mathbf{R}{ij}})\mathbf{g}_{j},
\end{split}
\label{eq:9}
\end{align}
where $\mathbf{R}_{ij} = \mathbf{T}^{H}_{j}\mathbf{M}^{\top}_{i}\mathbf{M}_{i}\mathbf{T}_{j}$ and $\overline{\mathbf{R}_{ij}}$ is conjugate to $\mathbf{R}_{ij}$.

The gradient of $\frac{\partial \mathbf{h}(\mathbf{g})}{\partial \mathbf{g}_{j}}$ can be written as
\begin{equation}
    \frac{\partial \mathbf{h}(\mathbf{g})}{\partial \mathbf{g}_{j}} = 
    \begin{bmatrix}
     \frac{\partial \mathbf{h}_{1}(\mathbf{g})}{\partial \mathbf{g}_{j}} & \frac{\partial \mathbf{h}_{2}(\mathbf{g})}{\partial \mathbf{g}_{j}} & ... & \frac{\partial \mathbf{h}_{N}(\mathbf{g})}{\partial \mathbf{g}_{j}} 
    \end{bmatrix} \in \mathbb{R}^{Ld \times N}
    \label{eq:grad_vector}
\end{equation}
Taking into consideration Eq.~\eqref{eq:grad_vector} and applying the chain rule, we derive the expression for $\frac{\partial \mathbf{h}(\mathbf{g})}{\partial \mathbf{g}_{j}}\mathbf{F}^{*}\mathbf{q}$ from Eq.~\eqref{eq:11}, where $\mathbf{q} \in \mathbb{R}^{N \times 1}$ is a real-valued vector, backpropagated from further layers. This way we can write
\begin{equation}
\frac{\partial \mathbf{h}(\mathbf{g})}{\partial \mathbf{g}_{j}}\mathbf{F}^{*}\mathbf{q} = \sum_{i=1}^{D}\frac{\partial \mathbf{h}_{i}(\mathbf{g})}{\partial \mathbf{g}_{j}}\mathbf{Q}_{i},\quad 
\mathbf{Q}_{i} = \mathbf{M}_{i}\mathbf{F}^{*}\mathbf{q}.
\label{eq:12}
\end{equation}
Let us derive the expression under the summation in Eq.~\eqref{eq:12} as
\begin{align}
&\frac{\partial \mathbf{h}_{i}(\mathbf{g})}{\partial \mathbf{g}_{j}}\mathbf{Q}_{i} = - e^{\alpha} \frac{\mathbf{M}_{i}\mathbf{\text{\boldmath$\lambda$}}^{*}_{\mathbf{K}} \odot \mathbf{Fy}}{(|\mathbf{M}_{i}\mathbf{\text{\boldmath$\lambda$}}_{\mathbf{K}}|^{2} + e^{\alpha}\sum_{d=1}^{D}|\mathbf{M}_{i}\mathbf{T}_{d}\mathbf{g}_{d}|^{2})^{2}} \cdot \nonumber \\
&\cdot(\mathbf{R}{ij} + \overline{\mathbf{R}{ij}})\mathbf{g}_{j}\mathbf{M}_{i}\mathbf{F}^{*}\mathbf{q} = - e^{\alpha}(\mathbf{R}{ij} + \overline{\mathbf{R}{ij}})\mathbf{g}_{j}\mathbf{M}_{i}\mathbf{z}.
\label{eq:13}
\vspace{-.15cm}
\end{align}
Here we use the notation $\mathbf{D_{K}} = diag(\mathbf{\text{\boldmath$\lambda$}}_{\mathbf{K}})$, $\mathbf{D_{G}}_{d} = diag(\mathbf{\text{\boldmath$\lambda$}}_{\mathbf{G}_{d}})$, and therefore $\mathbf{z} = \frac{\mathbf{D}^{*}_{\mathbf{K}}\mathbf{Fy}}{( |\mathbf{D}_{\mathbf{K}}|^{2} +  e^{\alpha}\sum_{d=1}^{D}|\mathbf{D}_{\mathbf{G}_d}|^{2})^{2}} \odot \mathbf{F}^{*}\mathbf{q}$. 

Here, the division operation corresponds to the element-wise division of the vector in the numerator by the corresponding diagonal elements of the diagonal matrix in the denominator. Here $diag(\mathbf{\mathbf{\text{\boldmath$\lambda$}}})$ denotes a square matrix with vector $\mathbf{\mathbf{\text{\boldmath$\lambda$}}}$ on the main diagonal and zeros otherwise.

Incorporating Eq.~\eqref{eq:13} into Eq.~\eqref{eq:12} and considering $\mathbf{Q} = \mathbf{F}^{*}\mathbf{q}$ we can write
\begin{align}
\begin{split}
\vspace{-.1cm}
\frac{\partial \mathbf{h}(\mathbf{g})}{\partial \mathbf{g}_{j}}\mathbf{Q} &= - e^{\alpha}\sum_{i=1}^{D}(\mathbf{R}_{ij} + \overline{\mathbf{R}_{ij}})\mathbf{g}_{j} \mathbf{M}_{i}\mathbf{z} = - e^{\alpha}\sum_{i=1}^{D}\left (\mathbf{T}^{H}_{j}\mathbf{M}^{\top}_{i}\mathbf{M}_{i}\mathbf{T}_{j} + \right. \\
& \left. + \overline{\mathbf{T}^{H}_{j}\mathbf{M}^{\top}_{i}\mathbf{M}_{i}\mathbf{T}_{j}}\right )\mathbf{g}_{j}\mathbf{z}_{i} = - e^{\alpha}\mathbf{P}^{\top}_{\mathbf{G}_{j}}\mathbf{S}^{\top}_{\mathbf{G}_{j}} \left (\mathbf{F}^{H}\sum_{i=1}^{D}\mathbf{M}^{\top}_{i}\mathbf{M}_{i}\mathbf{\text{\boldmath$\lambda$}}_{\mathbf{G}_{j}}\mathbf{z}_{i} + \right.\\
&+ \left. \mathbf{F}^{\top}\sum_{i=1}^{D}\mathbf{M}^{\top}_{i}\mathbf{M}_{i}\mathbf{\text{\boldmath$\lambda$}}^{*}_{\mathbf{G}_{j}}\mathbf{z}_{i} \right ) = - e^{\alpha}\mathbf{P}^{\top}_{\mathbf{G}_{j}}\mathbf{S}^{\top}_{\mathbf{G}_{j}} \left (\mathbf{F}^{H}(\mathbf{\mathbf{\text{\boldmath$\lambda$}}}_{\mathbf{G}_{j}} \odot \mathbf{z}) + \right.\\
&+ \left. \mathbf{F}^{\top}(\mathbf{\mathbf{\text{\boldmath$\lambda$}}}^{*}_{\mathbf{G}_{j}} \odot \mathbf{z}) \right ) = - e^{\alpha}\mathbf{P}^{\top}_{\mathbf{G}_{j}}\mathbf{S}^{\top}_{\mathbf{G}_{j}} \left (\mathbf{F}^{H}(\mathbf{\mathbf{\text{\boldmath$\lambda$}}}_{\mathbf{G}_{j}} \odot \mathbf{z}) + \overline{\mathbf{F}^{H}(\mathbf{\mathbf{\text{\boldmath$\lambda$}}}_{\mathbf{G}_{j}} \odot \mathbf{z}^{*}}) \right ).
\label{eq:long}
\end{split}
\end{align}
We note that $\mathbf{F}^{H}(\mathbf{\mathbf{\text{\boldmath$\lambda$}}}_{\mathbf{G}_{j}} \odot \mathbf{z}^{*})$ is real, therefore $\overline{\mathbf{F}^{H}(\mathbf{\mathbf{\text{\boldmath$\lambda$}}}_{\mathbf{G}_{j}} \odot \mathbf{z}^{*}}) = \mathbf{F}^{H}(\mathbf{\mathbf{\text{\boldmath$\lambda$}}}_{\mathbf{G}_{j}} \odot \mathbf{z}^{*})$. This way Eq.~\eqref{eq:long} can be rewritten as
\begin{align}
\begin{split}
\frac{\partial \mathbf{h}(\mathbf{g})}{\partial \mathbf{g}_{j}}\mathbf{Q}
&= - e^{\alpha}\mathbf{P}^{\top}_{\mathbf{G}_{j}}\mathbf{S}^{\top}_{\mathbf{G}_{j}} \left (\mathbf{F}^{H}(\mathbf{\mathbf{\text{\boldmath$\lambda$}}}_{\mathbf{G}_{j}} \odot \mathbf{z}) + \mathbf{F}^{H}(\mathbf{\mathbf{\text{\boldmath$\lambda$}}}_{\mathbf{G}_{j}} \odot \mathbf{z}^{*}) \right ) = \\
&= - e^{\alpha}\mathbf{P}^{\top}_{\mathbf{G}_{j}}\mathbf{S}^{\top}_{\mathbf{G}_{j}} \mathbf{F}^{H}\left (\mathbf{\mathbf{\text{\boldmath$\lambda$}}}_{\mathbf{G}_{j}} \odot (\mathbf{z} + \mathbf{z}^{*})\right ) = \\
&= - 2e^{\alpha}\mathbf{P}^{\top}_{\mathbf{G}_{j}}\mathbf{S}^{\top}_{\mathbf{G}_{j}} \mathbf{F}^{H}\left (\mathbf{\mathbf{\text{\boldmath$\lambda$}}}_{\mathbf{G}_{j}} \odot \text{Re}(\mathbf{z})\right ).
\label{eq:14}
\end{split}
\end{align}

This way the expression for the gradient of $f(\alpha, \mathbf{y}, \mathbf{g}_{d})$ with respect to the regularization kernel $\mathbf{g}_{d}$ is
\begin{align}
\begin{split}
&\frac{\partial f(\alpha, \mathbf{y}, \mathbf{g}_{d})}{\partial \mathbf{g}_{d}}\mathbf{q} = - 2e^{\alpha}\mathbf{P}^{\top}_{\mathbf{G}_{d}}\mathbf{S}^{\top}_{\mathbf{G}_{d}} \mathbf{F}^{H}\cdot \\
&\cdot \left [ \mathbf{\mathbf{\text{\boldmath$\lambda$}}}_{\mathbf{G}_{d}} \odot \text{Re} \left (\frac{\mathbf{D}^{*}_{\mathbf{K}}\mathbf{Fy}}{( |\mathbf{D}_{\mathbf{K}}|^{2} + e^{\alpha}\sum_{d=1}^{D}|\mathbf{D}_{\mathbf{G}_d}|^{2})^{2}} \odot \mathbf{F}^{*}\mathbf{q}\right )\right ].
\label{eq:fin}
\end{split}
\end{align}
The division operation in Eq.~\eqref{eq:fin} defines the element-wise division of the vector in the numerator by the corresponding diagonal elements of the diagonal matrix in the denominator.

\subsection{Gradient w.r.t. y}

Finally, we derive the expression for the gradient of $f(\alpha, \mathbf{y}, \mathbf{g}_{d})$ w.r.t. the input $\mathbf{y}$. Although this formula is not used in the training routine presented herein, we include the derivation to cover a general case of a larger pipeline which would comprise our network, e.g., when several Wiener filters are stacked into a sequence.
For this derivation, we use the original non FFT-based formulation of the Wiener filter:
\begin{equation}
f(\alpha, \mathbf{y}, \mathbf{g}_{d}) = (\mathbf{K}^{\top}\mathbf{K} + e^{\alpha}\sum_{d=1}^{D}\mathbf{G}_{d}^{\top}\mathbf{G}_{d})^{-1}\mathbf{K}^{\top}\mathbf{y}.
\label{eq:wiener}
\end{equation}
Here we denote $\mathbf{K}^{\top}\mathbf{K} + e^{\alpha}\sum_{d=1}^{D}\mathbf{G}_{d}^{\top}\mathbf{G}_{d} = \mathbf{B}$.
This way, Eq.~\eqref{eq:wiener} can be rewritten as $(\mathbf{K}^{\top}\mathbf{K} + e^{\alpha}\sum_{d=1}^{D}\mathbf{G}_{d}^{\top}\mathbf{G}_{d})^{-1}\mathbf{K}^{\top}\mathbf{y} = \mathbf{B}^{-1}\mathbf{K}^{\top}\mathbf{y}$, and the expression for the gradient of $f(\alpha, \mathbf{y}, \mathbf{g}_{d})$ w.r.t. the input $\mathbf{y}$ can be simplified as
\begin{equation}
\frac{\partial f(\alpha, \mathbf{y}, \mathbf{g}_{d})}{\partial \mathbf{y}} = \frac{\partial \mathbf{B}^{-1}\mathbf{K}^{\top} \mathbf{y}}{\partial \mathbf{y}} = \mathbf{K} \mathbf{B}^{-\top} =  \mathbf{K} \mathbf{B}^{-1}.
\label{eq:grad_y}
\end{equation}

As it was stated above, the matrices $\mathbf{K}$ and $\mathbf{G}_{d}$ are decomposed in the Fourier domain as
\begin{equation}
\mathbf{K} = \mathbf{F}^{H}\mathbf{D}_{\mathbf{K}}\mathbf{F},\ \mathbf{G}_{d} = \mathbf{F}^{H}\mathbf{D}_{\mathbf{G}_{d}}\mathbf{F},
\label{eq:circ}
\end{equation}
where $\mathbf{F} \in \mathbb{C}^{N \times N}$ is the Fourier (DFT) matrix, $\mathbf{F}^{H} \in \mathbb{C}^{N \times N}$ is its inverse, $\mathbf{D}_{\mathbf{K}}$, $\mathbf{D}_{\mathbf{G}_{d}} \in \mathbb{C}^{N \times N}$ are diagonal matrices.
Using the FFT-based inference of the Wiener filter and the notation defined in Eq.~\eqref{eq:denom}, the gradient of $f(\alpha, \mathbf{y}, \mathbf{g}_{d})$ w.r.t. the input $\mathbf{y}$ in Eq.~\eqref{eq:grad_y} can be rewritten as
\begin{equation}
    \frac{\partial f(\alpha, \mathbf{y}, \mathbf{g}_{d})}{\partial \mathbf{y}} = \mathbf{F}^{H}\mathbf{D}_{\mathbf{K}}({|\mathbf{D}_{\mathbf{K}}|^{2} + e^{\alpha}\sum_{d=1}^{D}|\mathbf{D_{G}}_{d}|^{2}})^{-1}\mathbf{F} = \mathbf{F}^{H}\mathbf{\Omega}^{-1}\mathbf{D}_{\mathbf{K}}\mathbf{F}.
\end{equation}

\section{Variation of kernels number and size for WF-K and WF-KPN}

The proposed models WF-K and WF-KPN employ a group of $D$ regularization kernels of size $K \times K$ to obtain the solution in the form of 
\begin{equation}\label{eq:wiener_fourier2}
\hat{\mathbf{x}} = \mathbf{F}^{H} \left ( \frac{\mathbf{D^{*}_{K}}\mathbf{F}\mathbf{y}}{|\mathbf{D_{K}}|^{2} + e^{\alpha}\sum_{d=1}^{D}|\mathbf{D_{G}}_{d}|^{2}} \right ).
\end{equation}
We performed an ablation study to understand the influence of the number and the size of the regularization kernels on the image restoration quality. Specifically, we trained both models, WF-K and WF-KPN, implementing $D=8$, $K=3$ and $D=24$, $K=5$ regularization kernels. For the ablation study, we used the same dataset and the training pipeline as in the main case. Namely, we implemented a Gaussian deblurring dataset, that was created by taking the ground-truth samples from the FMD~\cite{Zhang_2019_CVPR} and the dataset described in~\cite{AlKofahi2018ADL} and by cropping them into the tiles of size $256 \times 256$. 
All images were rescaled to the range [0,1]. 
During the training process, a ground-truth sample from 975 training samples is convolved with a randomly chosen blur kernel from a set of 25 training PSFs, followed by a perturbationn with i.i.d. Gaussian noise with standard deviation from the set (0.001, 0.005, 0.01, 0.05, 0.1). To evaluate the performance of the algorithms, we used 5 test sets of images distorted with the Gaussian noise of different levels. In particular, we used 230 ground-truth images and convolved them with a fixed blur kernel from a sub-set of 5 PSFs reserved for testing. Finally, the noisy observation of a blurry image is produced by adding the Gaussian noise, with standard deviations being taken from the set (0.001, 0.005, 0.01, 0.05, 0.1). Note that all ground-truth and the resulting distorted images are grayscale in all experiments (the false color Figures in the main paper were produced to emphasize fine details in the background).

Results in the Table~\ref{tab:ablation} show that increasing the number and the size of kernels in WF-K method leads to improvement in PSNR of nearly 0.15dB (in the low noise regime). Yet, at high noise levels, an increase of the size and the number of regularization kernels results in a marginal PSNR drop of 0.06dB.
For WF-KPN, increasing the number and the size of kernels improves the results for all noise levels, from nearly 0.4dB in the high noise regime to nearly 1.25dB at the low noise levels. Table~\ref{tab:ablation} shows that the increase of the parameters of the network expectedly leads to improvement of the results. In particular, increasing the number and the size of the kernels yields better performance at the low noise levels.

\begin{figure*}[t]
\captionsetup[subfigure]{justification=centering,labelformat=empty}
\captionsetup{format=plain, font=small, labelfont=bf}
  \centering
    \caption{Predicted kernels in WF-K and WF-KPN methods for different images and the corresponding restorations.} 
  \label{fig:kernels_output}
  \subfloat[Ground truth]{\includegraphics[width=0.13\textwidth,height=0.13\textwidth]{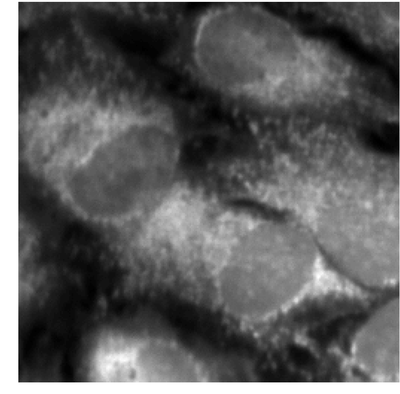}}
  \subfloat[Degraded]{\includegraphics[width=0.13\textwidth,height=0.13\textwidth]{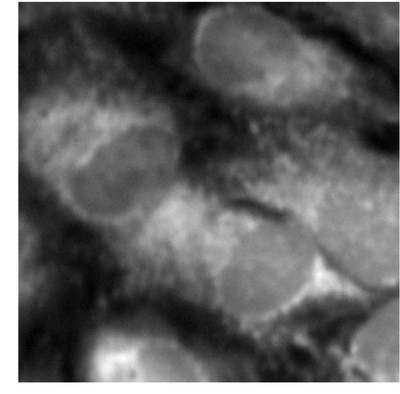}} 
  \subfloat[WF-K kernels]{\includegraphics[width=0.25\textwidth,height=0.133\textwidth]{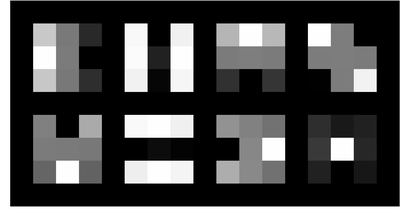}} 
  \subfloat[WF-KPN kernels]{\includegraphics[width=0.25\textwidth,height=0.133\textwidth]{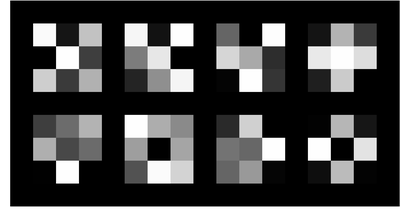}} 
  \subfloat[WF-K]{\includegraphics[width=0.13\textwidth,height=0.13\textwidth]{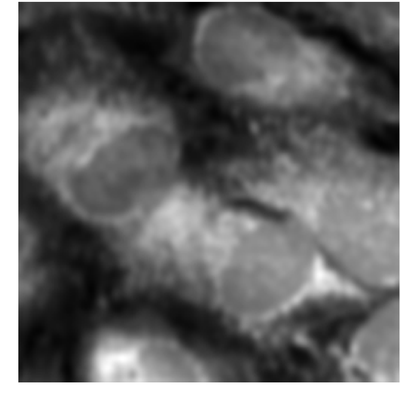}} 
  \subfloat[WF-KPN]{\includegraphics[width=0.13\textwidth,height=0.13\textwidth]{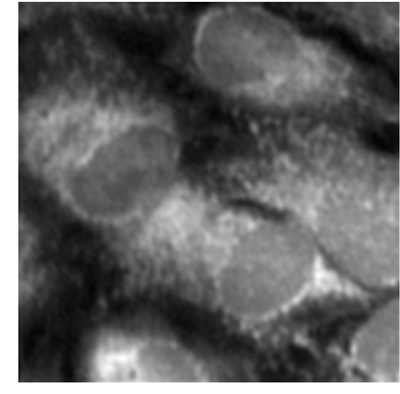}} \\
  \vspace{-.3cm}
  \subfloat{\includegraphics[width=0.13\textwidth,height=0.13\textwidth]{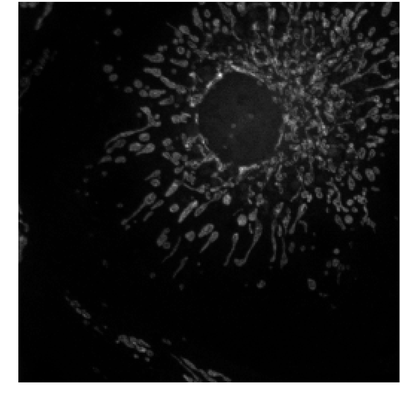}} 
  \subfloat{\includegraphics[width=0.13\textwidth,height=0.13\textwidth]{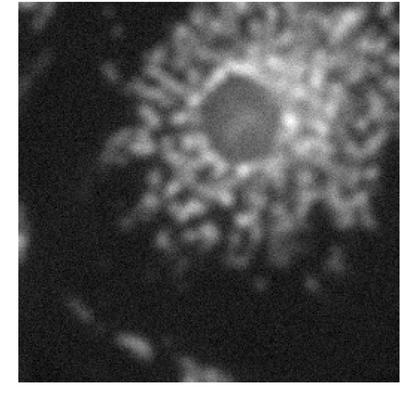}}
  \subfloat{\includegraphics[width=0.25\textwidth,height=0.133\textwidth]{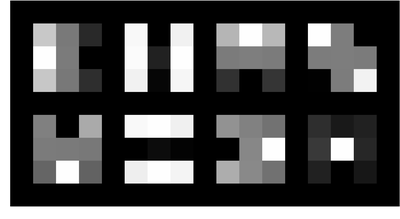}}
  \subfloat{\includegraphics[width=0.25\textwidth,height=0.133\textwidth]{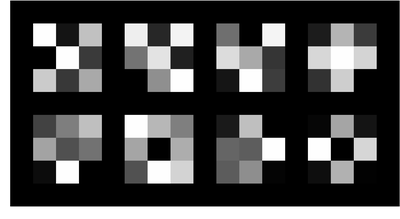}} 
  \subfloat{\includegraphics[width=0.13\textwidth,height=0.13\textwidth]{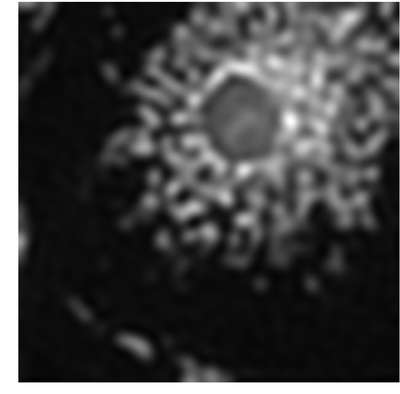}} 
  \subfloat{\includegraphics[width=0.13\textwidth,height=0.13\textwidth]{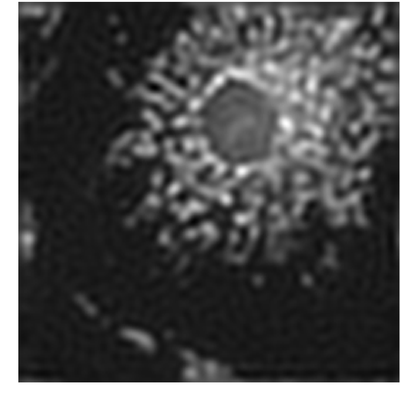}} \\
  \vspace{-.3cm}
  \subfloat{\includegraphics[width=0.13\textwidth,height=0.13\textwidth]{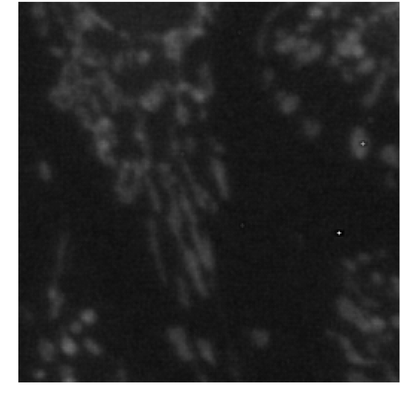}} 
  \subfloat{\includegraphics[width=0.13\textwidth,height=0.13\textwidth]{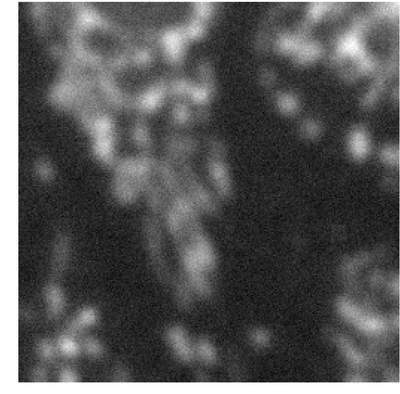}}
  \subfloat{\includegraphics[width=0.25\textwidth,height=0.133\textwidth]{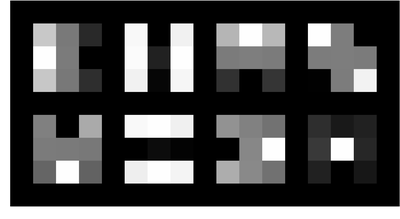}}
  \subfloat{\includegraphics[width=0.25\textwidth,height=0.133\textwidth]{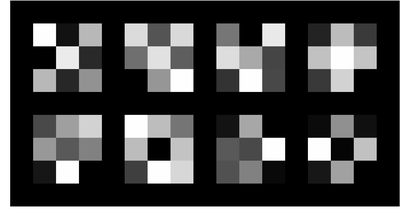}}
  \subfloat{\includegraphics[width=0.13\textwidth,height=0.13\textwidth]{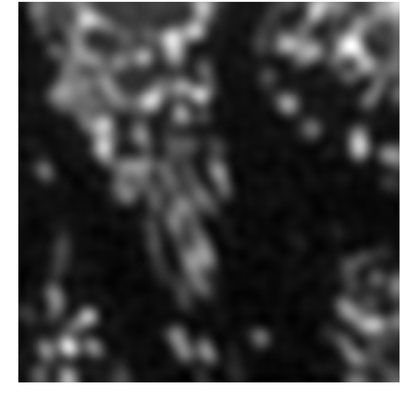}} 
  \subfloat{\includegraphics[width=0.13\textwidth,height=0.13\textwidth]{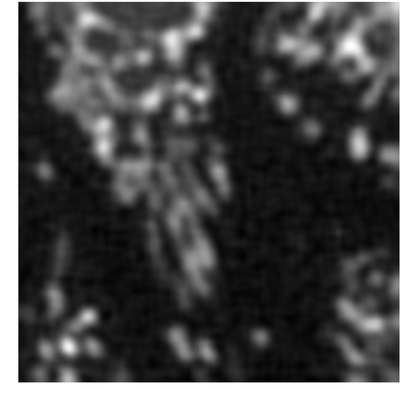}} \\
  \vspace{-.3cm}
  \subfloat{\includegraphics[width=0.13\textwidth,height=0.13\textwidth]{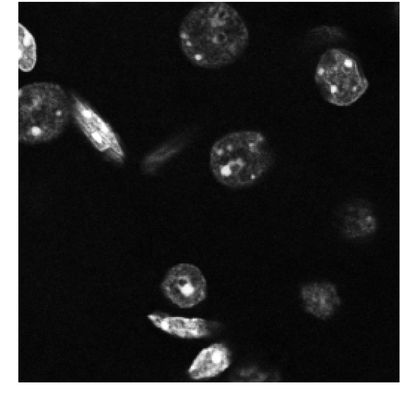}} 
  \subfloat{\includegraphics[width=0.13\textwidth,height=0.13\textwidth]{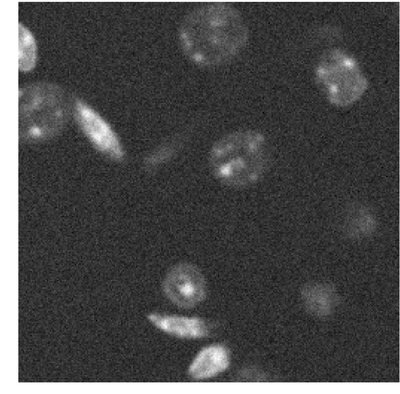}}
  \subfloat{\includegraphics[width=0.25\textwidth,height=0.133\textwidth]{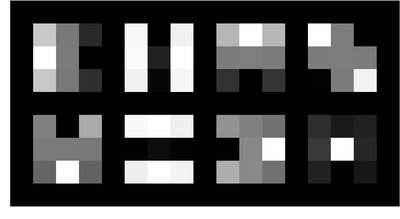}} 
  \subfloat{\includegraphics[width=0.25\textwidth,height=0.133\textwidth]{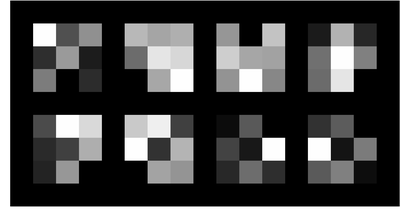}}
  \subfloat{\includegraphics[width=0.13\textwidth,height=0.13\textwidth]{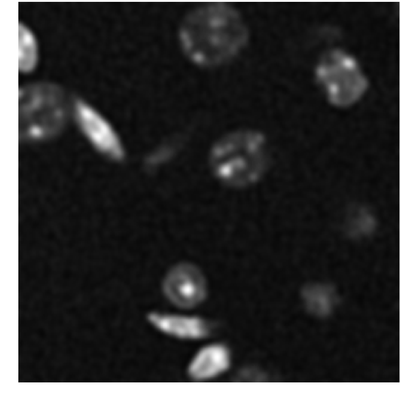}} 
  \subfloat{\includegraphics[width=0.13\textwidth,height=0.13\textwidth]{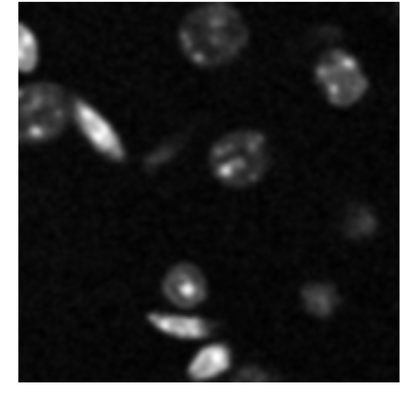}} \\
  \vspace{-.3cm}
  \subfloat{\includegraphics[width=0.13\textwidth,height=0.13\textwidth]{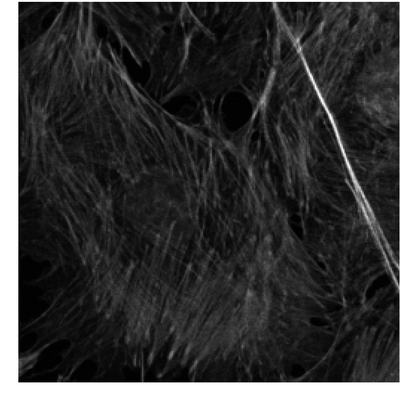}} 
  \subfloat{\includegraphics[width=0.13\textwidth,height=0.13\textwidth]{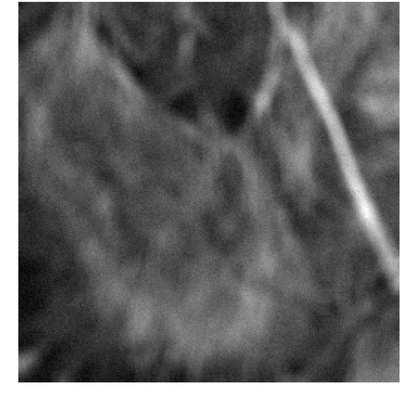}}
  \subfloat{\includegraphics[width=0.25\textwidth,height=0.133\textwidth]{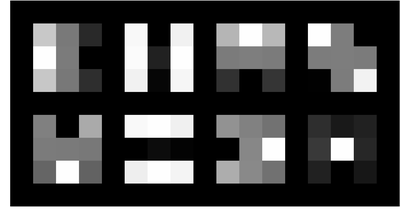}} 
  \subfloat{\includegraphics[width=0.25\textwidth,height=0.133\textwidth]{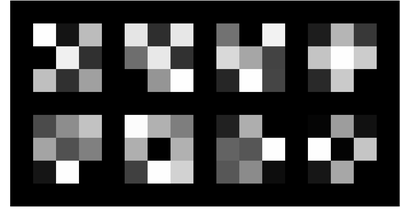}}
  \subfloat{\includegraphics[width=0.13\textwidth,height=0.13\textwidth]{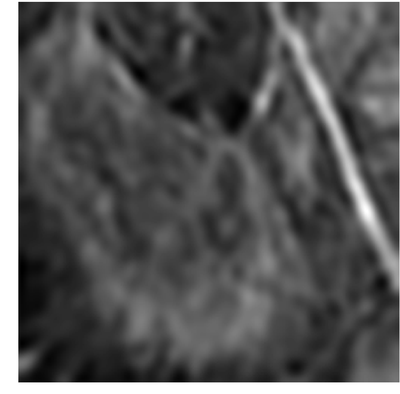}} 
  \subfloat{\includegraphics[width=0.13\textwidth,height=0.13\textwidth]{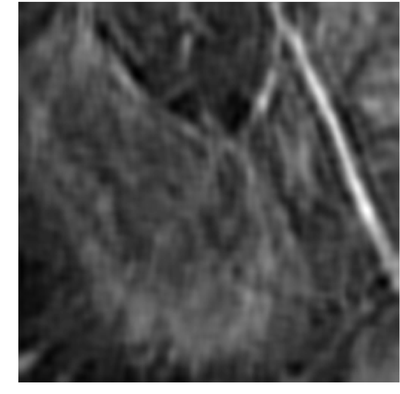}} \\
  \vspace{-.3cm}
  \subfloat{\includegraphics[width=0.13\textwidth,height=0.13\textwidth]{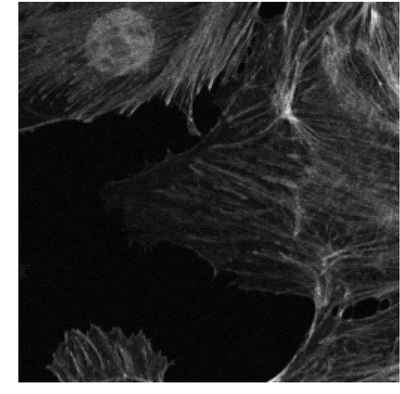}} 
  \subfloat{\includegraphics[width=0.13\textwidth,height=0.13\textwidth]{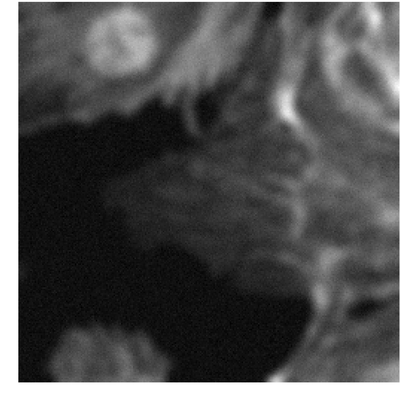}}
  \subfloat{\includegraphics[width=0.25\textwidth,height=0.133\textwidth]{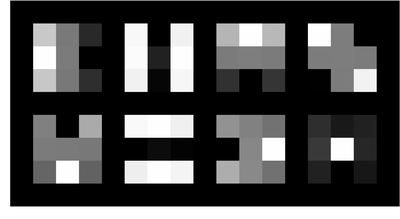}} 
  \subfloat{\includegraphics[width=0.25\textwidth,height=0.133\textwidth]{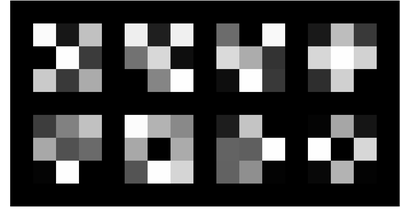}}
  \subfloat{\includegraphics[width=0.13\textwidth,height=0.13\textwidth]{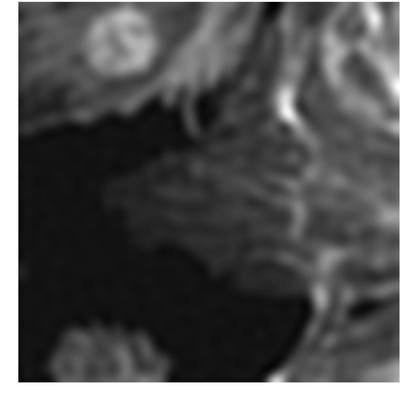}} 
  \subfloat{\includegraphics[width=0.13\textwidth,height=0.13\textwidth]{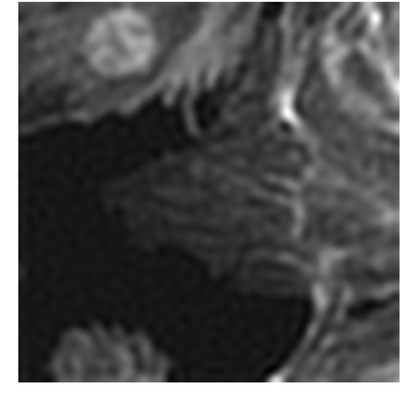}} \\
  \vspace{-.3cm}
  \subfloat{\includegraphics[width=0.13\textwidth,height=0.13\textwidth]{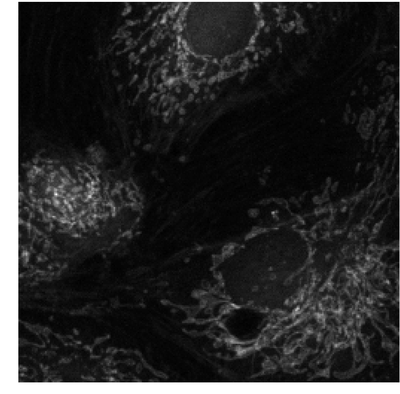}} 
  \subfloat{\includegraphics[width=0.13\textwidth,height=0.13\textwidth]{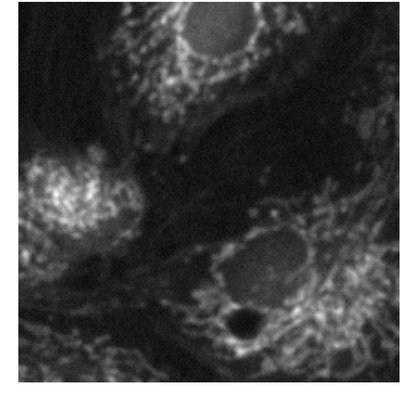}}
  \subfloat{\includegraphics[width=0.25\textwidth,height=0.133\textwidth]{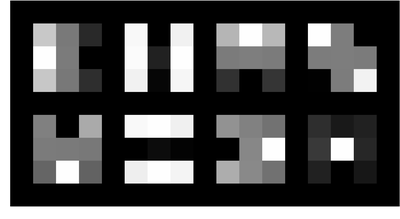}} 
  \subfloat{\includegraphics[width=0.25\textwidth,height=0.133\textwidth]{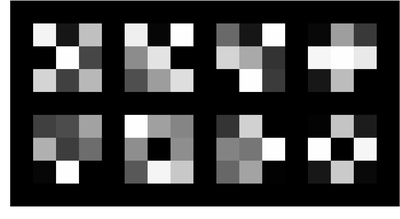}}
  \subfloat{\includegraphics[width=0.13\textwidth,height=0.13\textwidth]{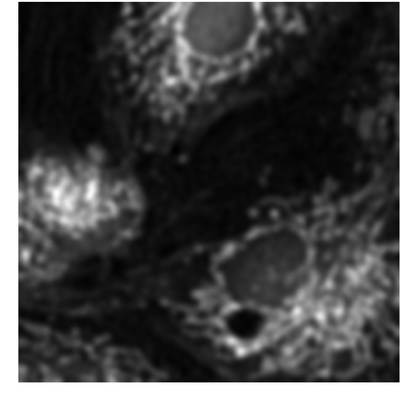}} 
  \subfloat{\includegraphics[width=0.13\textwidth,height=0.13\textwidth]{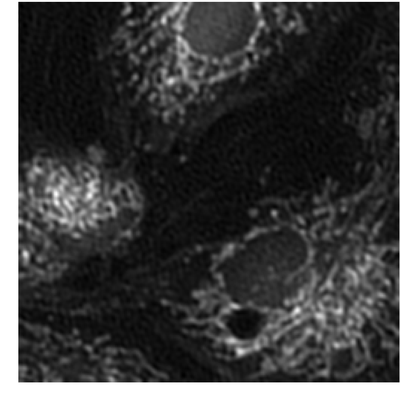}}
\end{figure*}

Regularization kernels, predicted with the models WF-K and WF-KPN with $D=8$, $K=3$, are shown in Figure \ref{fig:kernels_output} along with the results of the restoration with both models. 
The images show that WF-K provides a group of learnable kernels which are identical for all images, whereas WF-KPN predicts a group of regularization kernels per image. 
Results of the restoration with both models, presented in the Figure \ref{fig:kernels_output}, demonstrate that WF-KPN tends to restore images better than WF-K.

\begin{table*}[t]
\centering
\captionsetup{format=plain, font=small, labelfont=bf}
\caption{PSNR and SSIM comparisons on Gaussian image deblurring for five different noise levels for different number and size of the regularization kernels\label{tab:ablation}}
\vspace{.3cm}
\small
\resizebox{12cm}{!} {
\begin{tabular}{llllllllllll}
\specialrule{.1em}{.05em}{.05em} 
& \multicolumn{11}{c}{STD} \\
\specialrule{.1em}{.05em}{.05em}  
& & \multicolumn{2}{c}{0.001} & \multicolumn{2}{c}{0.005} & \multicolumn{2}{c}{0.01} & \multicolumn{2}{c}{0.05} & \multicolumn{2}{c}{0.1}\\
& & \multicolumn{1}{c}{PSNR} & \multicolumn{1}{c}{SSIM} & \multicolumn{1}{c}{PSNR} & \multicolumn{1}{c}{SSIM} & \multicolumn{1}{c}{PSNR} & \multicolumn{1}{c}{SSIM} & \multicolumn{1}{c}{PSNR} & \multicolumn{1}{c}{SSIM} & \multicolumn{1}{c}{PSNR} & \multicolumn{1}{c}{SSIM} \\ \hline
& \multicolumn{1}{|l|}{Input} & 36.23 & \multicolumn{1}{l|}{.8955} & 35.37 & \multicolumn{1}{l|}{.8791} & 33.93 & \multicolumn{1}{l|}{.8339} & 26.03 & \multicolumn{1}{l|}{.3858} & 21.14 & .1718 \\ \hline
\multirow{2}{*}{\textit{D} = 8, \textit{K} = 3} & \multicolumn{1}{|l|}{WF-K} & 35.66 & \multicolumn{1}{l|}{.8849} & 35.61 & \multicolumn{1}{l|}{.8834} & 35.45 & \multicolumn{1}{l|}{.8787} & 32.74 & \multicolumn{1}{l|}{.7950} & 29.27 & .6835 \\
&\multicolumn{1}{|l|}{WF-KPN} & 38.72 & \multicolumn{1}{l|}{.9253} & 37.98 & \multicolumn{1}{l|}{.9176} & 36.80 & \multicolumn{1}{l|}{.9028} & 32.33 & \multicolumn{1}{l|}{.8022} & 29.20 & .7259 \\ \hline
\multirow{2}{*}{\textit{D} = 24, \textit{K} = 5} & \multicolumn{1}{|l|}{WF-K} & 35.81 & \multicolumn{1}{l|}{.8861} & 35.75 & \multicolumn{1}{l|}{.8846} & 35.58 & \multicolumn{1}{l|}{.8798} & 32.75 & \multicolumn{1}{l|}{.7945} & 29.21 & .6807 \\
&\multicolumn{1}{|l|}{WF-KPN} & 39.95 & \multicolumn{1}{l|}{.9368} & 38.41 & \multicolumn{1}{l|}{.9218} & 37.20 & \multicolumn{1}{l|}{.9057} & 32.69 & \multicolumn{1}{l|}{.8095} & 29.20 & .7386 \\
\specialrule{.1em}{.05em}{.05em} 
\end{tabular}
}
\end{table*}

\section{Additional Results}
Additional results for deblurring of images corrupted with the Gaussian noise of different levels are presented in Fig.~\ref{fig:deblur_gaussian}. 
We observe that our methods are marginally inferior to the 
DMSP-NA and FDN on the lowest noise levels, however, our method WF-KPN-SA outperforms the other methods in the higher noise regimes, allowing the restoration of the finest image details.

\begin{figure*}[t]
  \captionsetup{format=plain, font=small, labelfont=bf}
  \centering
    \caption{Restoration of microscopy images scaled to the maximum intensity peaks (p) from the range (1, 2, 5, 10, 25, 50) and degraded by PSF and the Poisson noise. The metrics shown beneath each image are PSNR/SSIM.  The images are originally grayscale, but are shown in a pseudo-color to stress the details in different structures, textures, and in the background.} 
  \subfloat{\includegraphics[width=1.05\textwidth]{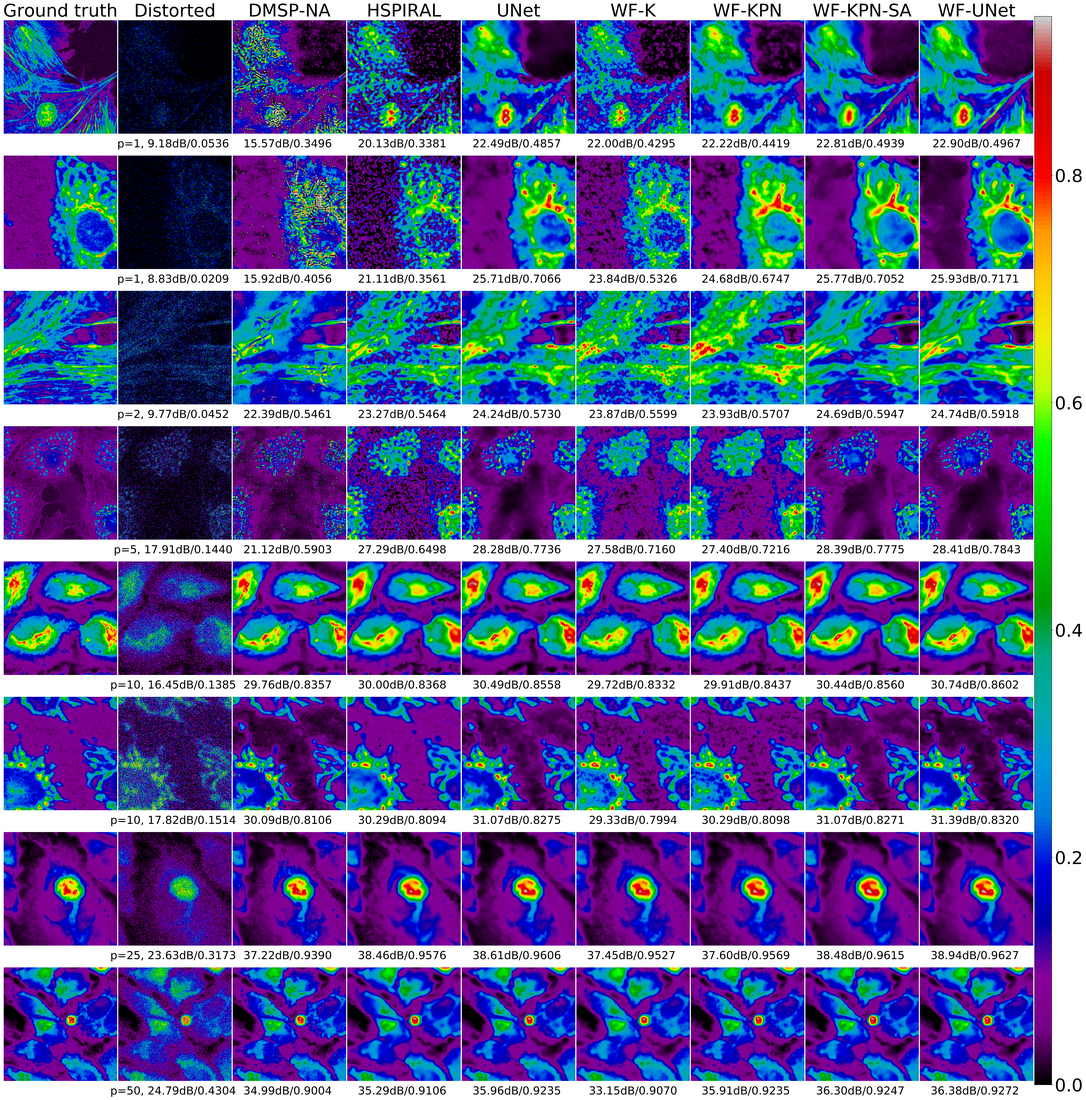}}
  \label{fig:deblur_poisson}
\end{figure*}

Additional study of the Poisson image deblurring, presented in Fig.~\ref{fig:deblur_poisson}, further proves that our methods WF-KPN-SA and WF-UNet allow to reconstruct the smallest image details with excellent values of the metrics.

\begin{figure*}[t]
  \captionsetup{format=plain, font=small, labelfont=bf}
  \centering
    \caption{Restoration of microscopy images degraded by PSF and Gaussian noise with the standard deviations ($\sigma$) from the set (0.001, 0.005, 0.01, 0.05, 0.1). The metrics shown beneath each image are PSNR/SSIM.  All images are originally grayscale, but are shown in pseudo-color to stress the details in different structures, textures, and in the background.} 
  \subfloat{\includegraphics[width=1.0\textwidth]{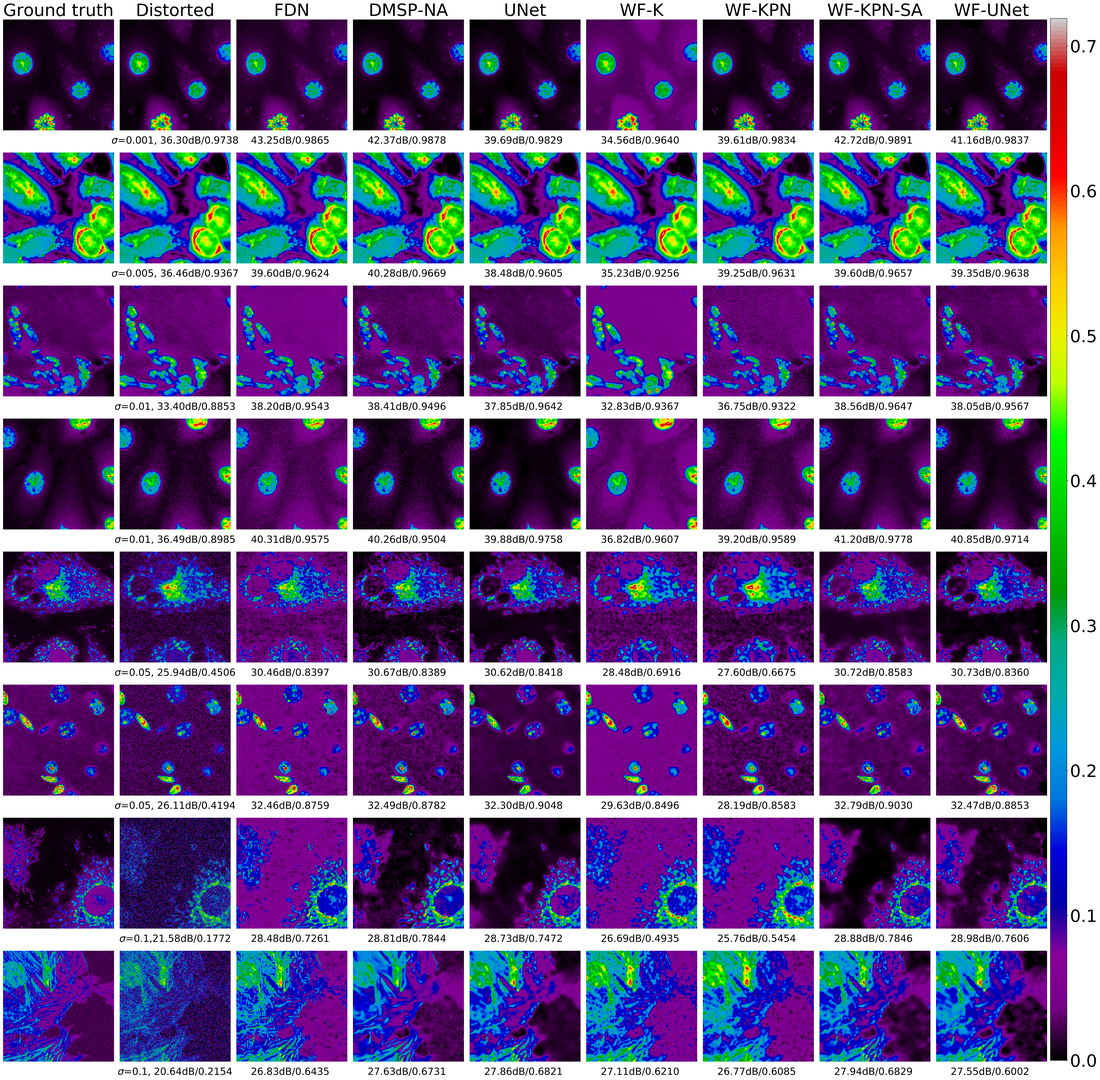}}
  \label{fig:deblur_gaussian}
\end{figure*}

\end{document}